\documentclass[10pt]{article}
\usepackage{graphicx}
\usepackage{amsmath}
\usepackage{amssymb}
\usepackage{caption2}
\begin{document}
\begin{center}
{\large {\bf \sc{  Analysis of the hidden-charm-hidden-strange tetraquark mass spectrum via the QCD sum rules }}} \\[2mm]
Zhi-Gang  Wang \footnote{E-mail: zgwang@aliyun.com.  }     \\
 Department of Physics, North China Electric Power University, Baoding 071003, P. R. China
\end{center}

\begin{abstract}
In the present work,  we construct the diquark-antidiquark type four-quark  currents to investigate  the  mass spectrum of the ground state hidden-charm-hidden-strange tetraquark states with the quantum numbers $J^{PC}=0^{++}$, $1^{+-}$, $1^{++}$ and $2^{++}$ via
the traditional QCD sum rules in a comprehensive way. We update old calculations,  perform new calculations and analysis in a rigorous way,  and  take account of the net  light-flavor $SU(3)$ breaking effects in a consistent way. And we make more reasonable identifications for the $X(3960)$, $X(4140)$, $X(4274)$, $X(4500)$, $X(4685)$ and $X(4700)$ and supersede some old identifications.
Furthermore, we consider our previous theoretical predictions, and make reasonable/suitable  identifications of the new LHCb states   $h_c(4000)$ and $\chi_{c1}(4010)$.
 \end{abstract}



\section{Introduction}
In 2003, the  Belle collaboration   observed   a narrow charmonium-like state $X(3872)$, the most enigmatical particle up to now, in the $\pi^+ \pi^- J/\psi$ invariant mass spectrum in the exclusive $B^\pm \to K^\pm \pi^+ \pi^- J/\psi$ decays \cite{X3872-2003}, which cannot be
suitably/comfortably accommodated in the traditional quark model as the $\chi_{c1}^\prime$ state with the spin-parity-charge-conjugation   $J^{PC}=1^{++}$ considering  the much smaller mass. Thereafter, a number of  charmonium-like states were observed experimentally \cite{PDG},   which cannot be accommodated suitably/comfortably in the traditional quark model owing to conflicts in one way or others, and were named as the $X$, $Y$ and $Z$ states, they are  very good  candidates for the exotic tetraquark states, molecular states, hybrid states, etc.  Among those exotic state candidates, the $X(4140)$ is a unique state and plays a role of benchmark.

In 2009,   the CDF collaboration observed the $X(4140)$ for the first time  in the $J/\psi\phi$ invariant  mass spectrum  in the exclusive $B^+ \to J/\psi\,\phi K^+$ decays  with a statistical
significance  larger than $3.8 \sigma$  \cite{CDF0903}.
 In 2011, the CDF collaboration confirmed the
$X(4140)$ in the exclusive $B^\pm\rightarrow J/\psi\,\phi K^\pm$ decays  with
 a statistical significance larger  than $5\sigma$, and  observed an evidence for the  $X(4274)$ in the $J/\psi\phi$ invariant  mass spectrum  with a statistical significance about   $3.1\sigma$
\cite{CDF1101}. Independent confirmation is still needed to really establish this exotic state.

In 2016, the LHCb collaboration performed the first full amplitude analysis of the exclusive $B^+\to J/\psi \phi K^+$ decays, and
 independently confirmed the two old particles $X(4140)$ and $X(4274)$ in the $J/\psi \phi$ invariant  mass spectrum  with statistical significances $8.4\sigma$ and $6.0\sigma$, respectively,  and
unambiguously established the spin-parity-charge-conjugation $J^{PC} =1^{++}$ with statistical significances $5.7\sigma$ and $5.8\sigma$, respectively \cite{LHCb-16061,LHCb-16062}. In addition, the LHCb collaboration observed  the two new particles $X(4500)$ and $X(4700)$ in the $J/\psi \phi$ invariant  mass spectrum  with statistical significances $6.1\sigma$ and $5.6\sigma$, respectively,  and established the corresponding quantum numbers $J^{PC} =0^{++}$ with statistical significances $4.0\sigma$ and $4.5\sigma$, respectively \cite{LHCb-16061,LHCb-16062}. The measured Breit-Wigner masses and widths are
\begin{flalign}
 & X(4140) : M = 4146.5 \pm 4.5 ^{+4.6}_{-2.8} \mbox{ MeV}\, , \, \Gamma = 83 \pm 21 ^{+21}_{-14} \mbox{ MeV} \, , \nonumber\\
 & X(4274) : M = 4273.3 \pm 8.3 ^{+17.2}_{-3.6} \mbox{ MeV}\, , \, \Gamma = 56 \pm 11 ^{+8}_{-11} \mbox{ MeV} \, ,\nonumber \\
 & X(4500) : M = 4506 \pm 11 ^{+12}_{-15} \mbox{ MeV} \, ,\, \Gamma = 92 \pm 21 ^{+21}_{-20} \mbox{ MeV} \, , \nonumber\\
 & X(4700) : M = 4704 \pm 10 ^{+14}_{-24} \mbox{ MeV} \, ,\, \Gamma = 120 \pm 31 ^{+42}_{-33} \mbox{ MeV} \, .
\end{flalign}

In 2021, the LHCb collaboration performed an improved full amplitude analysis of the exclusive $B^+\to J/\psi \phi K^+$ decays  and observed the tetraquark candidate  $X(4685)$  ($X(4630)$) in the $J/\psi \phi$ invariant mass spectrum  with a statistical significance of $15\sigma$  ($5.5\sigma$),  which has the favored  spin-parity identification $J^P=1^+$($1^-$), and the Breit-Wigner   mass and width   $4684 \pm 7 {}^{+13}_{-16}\,\rm{MeV}$  ($4626 \pm 16 {}^{+18}_{-110}\,\rm{MeV}$) and $126 \pm 15 {}^{+37}_{-41}\,\rm{MeV}$  ($174 \pm 27 {}^{+134}_{-73}\,\rm{MeV}$), respectively \cite{LHCb-X4685}. Furthermore, the
LHCb collaboration also observed two charged  tetraquark candidates $Z^+_{cs}(4000)$ and $Z^+_{cs}(4220)$ with explicit strangeness in the $J/\psi K^+$ invariant mass spectrum  with the preferred spin-parity $J^P=1^+$ \cite{LHCb-X4685}.

The $X(4140)$, $X(4274)$, $X(4500)$, $X(4630)$, $X(4685)$ and $X(4700)$ were observed in the $J/\psi\phi$ invariant mass spectrum, their quantum numbers  $J^{PC}=0^{++}$, $1^{++}$, $2^{++}$ for the S-wave couplings, and $0^{-+}$, $1^{-+}$, $2^{-+}$, $3^{-+}$ for the P-wave couplings. Considering the  $X$ states  are observed in the $J/\psi\phi$ invariant  mass spectrum, if their dominant Fock components are four-quark states, their
valence quark constituents are symbolically $\bar{c}c\bar{s}s$ rather than $\bar{c}c\bar{q}q$.
The LHCb collaboration established  the quantum numbers of the $X(4140)$ as  $J^{PC}=1^{++}$, which rules  out the
overwhelming   $D_s^*\bar{D}_s^*$ molecule identification with the assumption of the quantum numbers  $J^{PC}=0^{++}$ or $2^{++}$ after the discovery of the $X(4140)$ \cite{CDF0903,WZG-EPJC-Y4140-2009}, but does not rule  out the existence of the  $D_s^*\bar{D}_s^*$ molecular states with the $J^{PC}=0^{++}$ and $2^{++}$, those molecular states maybe observed experimentally in the future.

In this work, we will focus on the tetraquark scenario and the method of the traditional  QCD sum rules.
We always take the diquarks in color antitriplet  as the elementary building constituents  to analyze   the tetraquark states.   The diquarks  $\varepsilon^{abc}q^{T}_b C\Gamma q^{\prime}_c$  have  five independent spinor  structures, where the matrix structures $C\Gamma=C\gamma_5$, $C$, $C\gamma_\mu \gamma_5$,  $C\gamma_\mu $ and $C\sigma_{\mu\nu}$ (or $C\sigma_{\mu\nu}\gamma_5$) for the scalar ($S$), pseudoscalar ($P$), vector ($V$), axialvector ($A$)  and  tensor ($T$) diquarks, respectively, and the $a$, $b$, $c$ are color indexes. The tensor diquarks have both $J^P=1^+$ and $1^-$ Fock components, we
distinguish the spin-parity $J^P=1^+$ and $1^-$ components clearly, and name them as $\widetilde{A}$ and $\widetilde{V}$, respectively.

In Ref.\cite{X4140-tetraquark-Lebed}, Lebed  and Polosa identify   the  $X(3915)$ as the 1S $[cs]_{S}[\bar{c}\bar{s}]_{S}$
 tetraquark state with the $J^{PC}=0^{++}$ due to  lacking of the observed
decays to the final states $D\bar D$ and $D^*\bar{D}^*$, and attribute  the only known decay to the final state $J/\psi \omega$ to the $\omega-\phi$ mixing effects, and identify the $X(4140)$ as the $[cs]_{A}[\bar{c}\bar{s}]_{S}+[cs]_{S}[\bar{c}\bar{s}]_{A}$ tetraquark state with the $J^{PC}=1^{++}$   based on the effective  Hamiltonian with the spin-spin and spin-orbit  interactions, then in Ref.\cite{Maiani-X4140}, Maiani, Polosa and Riquer take the mass of the $X(4140)$ as input parameter, and obtain the mass spectrum of the $c\bar{c}s\bar{s}$ tetraquark states with positive parity, however, there is no room for the $X(4274)$, and they suggest the $X(4274)$
 corresponds to two, almost degenerate, unresolved lines with the $J^{PC}=0^{++}$ and $2^{++}$. In Ref.\cite{X4140-tetraquark-Stancu}, F. Stancu calculates  the mass spectrum of the $c\bar{c}s\bar{s}$ tetraquark states   via   a simple quark model with the chromomagnetic interaction but no correlated quarks, and obtain two lowest masses $4195\,\rm{MeV}$ and $4356\,\rm{MeV}$  with the $J^{PC}=1^{++}$. The value $4195\,\rm{MeV}$ is consistent with the LHCb data $4146.5 \pm 4.5 ^{+4.6}_{-2.8} \mbox{ MeV}$ \cite{LHCb-16061,LHCb-16062}.

In Ref.\cite{WangHuangTao-3900}, we  take the isospin limit and investigate the $S\bar{A}\pm A\bar{S}$ type hidden-charm tetraquark states with the $J^{PC}=1^{+\pm}$ respectively  via the traditional  QCD sum rules at length, and identify
  the $X(3872)$ and $Z_c(3900)$   as  the  diquark-antidiquark type tetraquark states with the quantum numbers  $J^{PC}=1^{++}$
and $1^{+-}$ respectively,
moreover, we investigate  the energy scale dependent behaviors of the QCD sum rules for the exotic states for the first time. Before Ref.\cite{WangHuangTao-3900},
the heavy quark masses and vacuum condensates were merely taken as input parameters at some mixed energy scales \cite{Nielsen-PRT}. In the subsequent works  \cite{WangTetraquarkCTP,Wang-tetra-formula,Wang-Huang-NPA-2014},  we
come up with an energy scale formula,
\begin{eqnarray}\label{ESF}
\mu&=&\sqrt{M^2_{X/Y/Z}-(2{\mathbb{M}}_Q)^2} \, ,
 \end{eqnarray}
 by introducing effective heavy quark masses ${\mathbb{M}}_Q$ to choose the most suitable   energy scales of the QCD spectral densities for  the hidden-charm (also hidden-bottom) tetraquark states. The energy scale formula is based on distinguishing  both the heavy and light degrees of freedom clearly,  can enhance the ground state contributions significantly (the pole contributions reach as large as $(40-60)\%$ while the central values exceed $50\%$) and can improve the convergent behaviors  of the operator product expansion significantly.

If we take the $X(4140)$ as the hidden-strange cousin of the $X(3872)$ with the symbolic quark structure $[sc]_S[\bar{s}\bar{c}]_A+[sc]_A[\bar{s}\bar{c}]_S$, then the mass difference  $M_{X(4140)}-M_{X(3872)}=275\,\rm{MeV}$, the light-flavor $SU(3)$ mass-breaking effect is about $m_s-m_q=135\,\rm{MeV}$, which is consistent with our naive expectation.

In Refs.\cite{Chen-Zhu-2011,Azizi1703}, the operator product expansion is carried out up to the vacuum condensates of dimension 8 to study the $[sc]_S[\bar{s}\bar{c}]_A+[sc]_A[\bar{s}\bar{c}]_S$  tetraquark state with the $J^{PC}=1^{++}$, Chen and Zhu obtain the tetraquark  mass $4.07\pm 0.10\,\rm{GeV}$ \cite{Chen-Zhu-2011}, Agaev, Azizi, and  Sundu obtain the tetraquark  mass $4.183 \pm 0.115\,\rm{GeV}$ \cite{Azizi1703}, which are all compatible with the experimental data $4146.5\pm 3.0\,\rm{MeV}$ from Particle Data Group within uncertainties \cite{PDG}, however, the pole contributions $44.4\%$ and $23\%$ in Ref.\cite{Chen-Zhu-2011} and Ref.\cite{Azizi1703} respectively are failed to satisfy the pole dominance criterion.

In Ref.\cite{Wang1607-Y4140}, we explore the $J^{PC}=1^{+\pm}$ tetraquark states with the symbolic diquark structures $[sc]_S[\bar{s}\bar{c}]_A\pm[sc]_A[\bar{s}\bar{c}]_S$ and $[sc]_P[\bar{s}\bar{c}]_V\mp[sc]_V[\bar{s}\bar{c}]_P$, respectively. We carry out the operator product expansion up to   the vacuum condensates of dimension 10, take  the energy scale formula to choose suitable energy scales and  investigate the ground states  at length, the predictions do not favor  identifying the $X(4140)$ as the $J^{PC}=1^{++}$  tetraquark state. However, we should bear in mind that in Ref.\cite{Wang1607-Y4140}, we do not apply the modified energy scale formula,
\begin{eqnarray}\label{MESF}
\mu&=&\sqrt{M^2_{X/Y/Z}-(2{\mathbb{M}}_Q)^2}-k{\mathbb{M}}_s \, ,
 \end{eqnarray}
to take the light flavor $SU(3)$ breaking effects into account due to the effective $s$-quark mass ${\mathbb{M}}_s$, and do not apply  the equation  $t^a_{ij}t^a_{mn}=-\frac{1}{6}\delta_{ij}\delta_{mn}+
\frac{1}{2}\delta_{jm}\delta_{in}$ in the color space to calculate the higher dimensional vacuum condensates rigorously but resort to some approximations, where $t^a=\frac{\lambda^a}{2}$, the $k$ is the number of the valence $s$-quark. Updated analysis is still needed.

In Ref.\cite{WangZG-Di-Y4140}, we apply both  the $[sc]_{\tilde{A}}[\bar{s}\bar{c}]_A+[sc]_A[\bar{s}\bar{c}]_{\tilde{A}}$ and $[sc]_{\tilde{V}}[\bar{s}\bar{c}]_V-[sc]_V[\bar{s}\bar{c}]_{\tilde{V}}$ type currents with the quantum numbers $J^{PC}=1^{++}$ to investigate  the mass of the $X(4140)$ at length. The predictions favor identifying the $X(4140)$ as the $[sc]_{\tilde{V}}[\bar{s}\bar{c}]_V-[sc]_V[\bar{s}\bar{c}]_{\tilde{V}}$  tetraquark state. Then we calculate the hadronic coupling constant $g_{XJ/\psi\phi}$ based on  rigorous  quark-hadron duality, and get the partial decay width $\Gamma(X(4140)\to J/\psi \phi)=86.9\pm22.6\,{\rm{MeV}}$, which is in very good  agreement with the experimental data $83\pm 21^{+21}_{-14} {\mbox{ MeV}}$ from the LHCb collaboration \cite{LHCb-16061,LHCb-16062}. However, we also do not apply the modified energy scale formula
to take the light flavor  $SU(3)$ breaking effects into account, see Eq.\eqref{MESF}.

In Ref.\cite{WZG-Y4274-octet}, we tentatively identify the $X(4274)$ as the color octet-octet  type tetraquark state (or molecule-like state) with the $J^{PC}=1^{++}$, and apply  the   $i\gamma_5\lambda^a \otimes \gamma_\mu \lambda^a$-type  current to explore its mass and width at length,  the predicted mass (width) favors (disfavors) identifying  the  $X(4274)$  as the color octet-octet  type  tetraquark  state (strongly). If the $X(4274)$ is the traditional charmonium  state $\chi_{c1}(\rm 3P)$,  it is important to explore the two-body strong decay $X(4274)\to J/\psi \omega$ experimentally to diagnose its nature, and we suggest the LHCb and other collaborations could perform such measurements.
In Ref.\cite{WZG-X4274-APPB}, we apply  the $[sc]_A[\bar{s}\bar{c}]_V-[sc]_V[\bar{s}\bar{c}]_A$ type tensor  current to explore the mass and width of the $X(4274)$ at length, the predictions  favor identifying the $X(4274)$ as the   $[sc]_A[\bar{s}\bar{c}]_V-[sc]_V[\bar{s}\bar{c}]_A$   tetraquark state with a relative P-wave between the diquark and antidiquark pairs. It is a possible identification of the $X(4274)$, which does not suffer from shortcomings in sense of treating scheme.

In Ref.\cite{WZG-X4140-X4685}, we  extend our  previous works \cite{WangZG-Di-Y4140} to investigate   the new tetraquark candidate $X(4685)$ as the first radial excitation  of the $X(4140)$, and acquire the mass $M_{X}=4.70\pm0.12\,\rm{GeV}$, which is in very good agreement with  the experimental data $4684 \pm 7 {}^{+13}_{-16}\,\rm{MeV}$ from the LHCb collaboration \cite{LHCb-X4685}. However, we neglect  the light flavor  $SU(3)$ breaking effects in applying the energy scale formula.

In Refs.\cite{X3915-X4500-EPJA-WZG,X3915-X4500-EPJC-WZG}, we tentatively identify the  $X(3915)$ and $X(4500)$ as the 1S and 2S  $[sc]_{A}[\bar{s}\bar{c}]_A$ tetraquark states with the $J^{PC}=0^{++}$, respectively,  and investigate them at length, and  reproduce the experimental values of the masses. The inclusion of the  2S state beyond the 1S state in the QCD sum rules leads to smaller 1S state mass \cite{Wang1502-Y4140}, which happens to lie in the energy region of the $X(3915)$. If only the 1S states are included, the  masses of the 1S state  $[sc]_{A}[\bar{s}\bar{c}]_A$ tetraquark states with the $J^{PC}=0^{++}$ and $2^{++}$ from the QCD sum rules are about $3.98\pm0.08\,\rm{GeV}$ and $4.13\pm0.08\,\rm{GeV}$, respectively \cite{Wang1502-Y4140}. In Refs.\cite{X3915-X4500-EPJC-WZG,Wang1502-Y4140}, we also neglect  the light flavor  $SU(3)$ breaking effects in applying the energy scale formula. In a short summary, there exist shortcomings in one way or the other in the old analysis \cite{Chen-Zhu-2011,Azizi1703,Wang1607-Y4140,
WangZG-Di-Y4140,WZG-X4140-X4685,X3915-X4500-EPJC-WZG,Wang1502-Y4140}.

There also exist other interpretations for those $X$ states, irrespective of the tetraquark scenario \cite{X4140-X4700-tetra-LuQF,X4140-X4700-tetra-ChenHX,X4140-X4700-tetra-Ferretti-SB,
X4140-X4700-tetra-WuJ,X4140-X4700-tetra-HuangF,X4140-X4700-tetra-HuangHX},   molecule scenario \cite{X4140-X4700-mole-WangE,X4140-X4700-mole-PengFZ}, or traditional charmonium scenario \cite{X4140-X4700-cc-LiuXH,X4140-X4700-cc-ChenDY,X4140-X4700-cc-ZhongXH}, which have both advantages and shortcomings, no definite conclusion can be obtained.

In 2022,  the LHCb collaboration observed the $X(3960)$ in the $D_s^+D_s^-$ invariant mass spectrum   with the significance of $12.6\,\sigma$ in the exclusive $B^{+} \to D^{+}_s D^{-}_s K^{+}$ decays, and  the identification of the quantum numbers $J^{PC}=0^{++}$ is  favored \cite{LHCb3960-2022}. The measured Breit-Wigner mass and width are $M = 3956 \pm 5\pm 10 $ MeV and $\Gamma = 43\pm 13\pm 8 $ MeV, respectively. Compared with the $X(3912)$, it is more reasonable to identify  the $X(3960)$ as the ground state $cs\bar{c}\bar{s}$ tetraquark state.

The modified energy scale formula
$\mu=\sqrt{M^2_{X/Y/Z}-(2{\mathbb{M}}_Q)^2}-k{\mathbb{M}}_s $ plays an important role in exploring the  light flavor $SU(3)$ breaking effects in the multiquark states \cite{WZG-IJMPA-Pcs4469,WZG-mole-IJMPA,WZG-XQ-mole-penta,WZG-XQ-mole-EPJA,
WZG-tetra-psedo-NPB,WZG-Zcs3985-4110,WZG-NPB-cucd,WZG-NPB-cscs}, we should take it into account. Although sometimes the traditional charmonium scenario also works  in interpreting the higher charmonium states \cite{cc-Godfrey}, it is very interesting to explore the tetraquark scenario and molecule scenario.    We  have performed comprehensive and consistent  investigations for  the hidden-charm tetraquark states with the quantum numbers $J^{PC}=0^{++}$,  $0^{-+}$, $0^{--}$, $1^{--}$, $1^{-+}$, $1^{+-}$, $2^{++}$ \cite{WZG-tetra-psedo-NPB,WZG-HC-spectrum-PRD,WZG-EPJC-P-wave,WZG-EPJC-P-2P}, hidden-bottom tetraquark states with the quantum numbers $J^{PC}=0^{++}$, $1^{+-}$, $2^{++}$ \cite{WZG-HB-spectrum-EPJC}, hidden-charm molecular states with the quantum numbers $J^{PC}=0^{++}$, $1^{+-}$, $2^{++}$ \cite{WZG-mole-IJMPA}, doubly-charm tetraquark (molecular) states with the quantum numbers $J^{P}=0^{+}$, $1^{+}$, $2^{+}$ \cite{WZG-tetra-cc-EPJC} (\cite{WZG-XQ-mole-EPJA}), hidden-charm pentaquark (molecular) states with the quantum numbers $J^{P}={\frac{1}{2}}^{-}$, ${\frac{3}{2}}^{-}$, ${\frac{5}{2}}^{-}$ \cite{WZG-penta-cc-IJMPA-2050003}(\cite{XWWang-penta-mole}), and make reasonable/suitable  identifications of the existing exotic states.

Now we extend our previous works to investigate the mass spectrum of the $cs\bar{c}\bar{s}$ tetraquark states with the quantum numbers $J^{PC}=0^{++}$, $1^{+-}$, $1^{++}$ and $2^{++}$ in a comprehensive and consistent way, and we accomplish the operator product expansion up to the vacuum condensates of dimension 10 rigorously (satisfying the counting rules) with the formula $t^a_{ij}t^a_{mn}=-\frac{1}{6}\delta_{ij}\delta_{mn}+\frac{1}{2}\delta_{jm}\delta_{in}$, and apply the modified energy scale formula $\mu=\sqrt{M^2_{X/Y/Z}-(2{\mathbb{M}}_Q)^2}-2{\mathbb{M}}_s$ to account for the
light-flavor $SU(3)$ breaking effects, and revisit identifications of the $X$ states.

The article is arranged as follows:  we acquire the QCD sum rules for the  hidden-charm-hidden-strange tetraquark states in section 2; in section 3, we   present the numerical results and discussions; section 4 is reserved for our conclusion.

\section{QCD sum rules for  the  hidden-charm-hidden-strange  tetraquark states}
At the beginning, we write down  the two-point correlation functions $\Pi(p)$, $\Pi_{\mu\nu}(p)$ and $\Pi_{\mu\nu\alpha\beta}(p)$,
\begin{eqnarray}\label{CF-Pi}
\Pi(p)&=&i\int d^4x e^{ip \cdot x} \langle0|T\Big\{J(x)J^{\dagger}(0)\Big\}|0\rangle \, ,\nonumber\\
\Pi_{\mu\nu}(p)&=&i\int d^4x e^{ip \cdot x} \langle0|T\Big\{J_\mu(x)J_{\nu}^{\dagger}(0)\Big\}|0\rangle \, ,\nonumber\\
\Pi_{\mu\nu\alpha\beta}(p)&=&i\int d^4x e^{ip \cdot x} \langle0|T\Big\{J_{\mu\nu}(x)J_{\alpha\beta}^{\dagger}(0)\Big\}|0\rangle \, ,
\end{eqnarray}
where the diquark-antidiquark type currents,
\begin{eqnarray}
J(x)&=&J_{SS}(x)\, , \, J_{AA}(x)\, , \, J_{\widetilde{A}\widetilde{A}}(x)\, , \, J_{VV}(x)\, , \, J_{\widetilde{V}\widetilde{V}}(x)\, , \, J_{PP}(x)\, , \nonumber\\
J_\mu(x)&=&J^{SA}_{-,\mu}(x)\, , \, J_{-,\mu}^{\widetilde{A}A}(x)\, , \, J_{-,\mu}^{\widetilde{V}V}(x)\, , \, J^{PV}_{-,\mu}(x)\, , \, J^{SA}_{+,\mu}(x)\, , \, J_{+,\mu}^{\widetilde{V}V}(x)\, , \, J_{+,\mu}^{\widetilde{A}A}(x)\, , \, J^{PV}_{+,\mu}(x)\, , \nonumber\\
J_{\mu\nu}(x)&=&J^{AA}_{-,\mu\nu}(x)\, , \,J^{S\widetilde{A}}_{-,\mu\nu}(x)\, , \,J^{VV}_{-,\mu\nu}(x)\, , \, J^{S\widetilde{A}}_{+,\mu\nu}(x)\, , \, J^{AA}_{+,\mu\nu}(x)\, , \, J^{VV}_{+,\mu\nu}(x)\, ,
\end{eqnarray}
and
\begin{eqnarray}
J_{SS}(x)&=&\varepsilon^{ijk}\varepsilon^{imn}s^{T}_j(x)C\gamma_5 c_k(x)  \bar{s}_m(x)\gamma_5 C \bar{c}^{T}_n(x) \, ,\nonumber \\
J_{AA}(x)&=&\varepsilon^{ijk}\varepsilon^{imn}s^{T}_j(x)C\gamma_\mu c_k(x)  \bar{s}_m(x)\gamma^\mu C \bar{c}^{T}_n(x) \, ,\nonumber \\
J_{\tilde{A}\tilde{A}}(x)&=&\varepsilon^{ijk}\varepsilon^{imn}s^{T}_j(x)C\sigma^v_{\mu\nu} c_k(x)  \bar{s}_m(x)\sigma_v^{\mu\nu} C \bar{c}^{T}_n(x) \, ,\nonumber \\
J_{VV}(x)&=&\varepsilon^{ijk}\varepsilon^{imn}s^{T}_j(x)C\gamma_\mu\gamma_5 c_k(x)  \bar{s}_m(x)\gamma_5\gamma^\mu C \bar{c}^{T}_n(x) \, ,\nonumber \\
J_{\tilde{V}\tilde{V}}(x)&=&\varepsilon^{ijk}\varepsilon^{imn}s^{T}_j(x)C\sigma^t_{\mu\nu} c_k(x)  \bar{s}_m(x)\sigma_t^{\mu\nu} C \bar{c}^{T}_n(x) \, ,\nonumber \\
J_{PP}(x)&=&\varepsilon^{ijk}\varepsilon^{imn}s^{T}_j(x)Cc_k(x)  \bar{s}_m(x) C \bar{c}^{T}_n(x) \, ,
\end{eqnarray}

\begin{eqnarray}
J^{SA}_{-,\mu}(x)&=&\frac{\varepsilon^{ijk}\varepsilon^{imn}}{\sqrt{2}}
\Big[s^{T}_j(x)C\gamma_5c_k(x) \bar{s}_m(x)\gamma_\mu C \bar{c}^{T}_n(x)-s^{T}_j(x)C\gamma_\mu c_k(x)\bar{s}_m(x)\gamma_5C \bar{c}^{T}_n(x) \Big] \, ,\nonumber\\
J^{AA}_{-,\mu\nu}(x)&=&\frac{\varepsilon^{ijk}\varepsilon^{imn}}{\sqrt{2}}
\Big[s^{T}_j(x) C\gamma_\mu c_k(x) \bar{s}_m(x) \gamma_\nu C \bar{c}^{T}_n(x)  -s^{T}_j(x) C\gamma_\nu c_k(x) \bar{s}_m(x) \gamma_\mu C \bar{c}^{T}_n(x) \Big] \, , \nonumber\\
J^{S\widetilde{A}}_{-,\mu\nu}(x)&=&\frac{\varepsilon^{ijk}\varepsilon^{imn}}{\sqrt{2}}
\Big[s^{T}_j(x)C\gamma_5 c_k(x)  \bar{s}_m(x)\sigma_{\mu\nu} C \bar{c}^{T}_n(x)- s^{T}_j(x)C\sigma_{\mu\nu} c_k(x)  \bar{s}_m(x)\gamma_5 C \bar{c}^{T}_n(x) \Big] \, , \nonumber\\
J_{-,\mu}^{\widetilde{A}A}(x)&=&\frac{\varepsilon^{ijk}\varepsilon^{imn}}{\sqrt{2}}
\Big[s^{T}_j(x)C\sigma_{\mu\nu}\gamma_5 c_k(x)\bar{s}_m(x)\gamma^\nu C \bar{c}^{T}_n(x)-s^{T}_j(x)C\gamma^\nu c_k(x)\bar{s}_m(x)\gamma_5\sigma_{\mu\nu} C \bar{c}^{T}_n(x) \Big] \, , \nonumber\\
J_{-,\mu}^{\widetilde{V}V}(x)&=&\frac{\varepsilon^{ijk}\varepsilon^{imn}}{\sqrt{2}}
\left[s^{T}_j(x)C\sigma_{\mu\nu} c_k(x)\bar{s}_m(x)\gamma_5\gamma^\nu C \bar{c}^{T}_n(x)+s^{T}_j(x)C\gamma^\nu \gamma_5c_k(x)\bar{s}_m(x) \sigma_{\mu\nu} C \bar{c}^{T}_n(x) \right] \, , \nonumber\\
J^{VV}_{-,\mu\nu}(x)&=&\frac{\varepsilon^{ijk}\varepsilon^{imn}}{\sqrt{2}}
\Big[s^{T}_j(x) C\gamma_\mu \gamma_5c_k(x) \bar{s}_m(x) \gamma_5\gamma_\nu C \bar{c}^{T}_n(x)  -s^{T}_j(x) C\gamma_\nu\gamma_5 c_k(x) \bar{s}_m(x) \gamma_5\gamma_\mu C \bar{c}^{T}_n(x) \Big] \, , \nonumber\\
J^{PV}_{-,\mu}(x)&=&\frac{\varepsilon^{ijk}\varepsilon^{imn}}{\sqrt{2}}
\Big[s^{T}_j(x)Cc_k(x) \bar{s}_m(x)\gamma_5\gamma_\mu C \bar{c}^{T}_n(x)+s^{T}_j(x)C\gamma_\mu \gamma_5c_k(x)\bar{s}_m(x)C \bar{c}^{T}_n(x) \Big] \, ,
\end{eqnarray}

\begin{eqnarray}
J^{SA}_{+,\mu}(x)&=&\frac{\varepsilon^{ijk}\varepsilon^{imn}}{\sqrt{2}}
\Big[s^{T}_j(x)C\gamma_5c_k(x) \bar{s}_m(x)\gamma_\mu C \bar{c}^{T}_n(x)+s^{T}_j(x)C\gamma_\mu c_k(x)\bar{s}_m(x)\gamma_5C \bar{c}^{T}_n(x) \Big] \, ,\nonumber\\
J^{S\widetilde{A}}_{+,\mu\nu}(x)&=&\frac{\varepsilon^{ijk}\varepsilon^{imn}}{\sqrt{2}}
\Big[s^{T}_j(x)C\gamma_5 c_k(x)  \bar{s}_m(x)\sigma_{\mu\nu} C \bar{c}^{T}_n(x)+ s^{T}_j(x)C\sigma_{\mu\nu} c_k(x)  \bar{s}_m(x)\gamma_5 C \bar{c}^{T}_n(x) \Big] \, , \nonumber\\
J_{+,\mu}^{\widetilde{V}V}(x)&=&\frac{\varepsilon^{ijk}\varepsilon^{imn}}{\sqrt{2}}
\left[s^{T}_j(x)C\sigma_{\mu\nu} c_k(x)\bar{s}_m(x)\gamma_5\gamma^\nu C \bar{c}^{T}_n(x)-s^{T}_j(x)C\gamma^\nu \gamma_5c_k(x)\bar{s}_m(x) \sigma_{\mu\nu} C \bar{c}^{T}_n(x) \right] \, , \nonumber\\
J_{+,\mu}^{\widetilde{A}A}(x)&=&\frac{\varepsilon^{ijk}\varepsilon^{imn}}{\sqrt{2}}
\left[s^{T}_j(x)C\sigma_{\mu\nu}\gamma_5 c_k(x)\bar{s}_m(x)\gamma^\nu C \bar{c}^{T}_n(x)+s^{T}_j(x)C\gamma^\nu c_k(x)\bar{s}_m(x)\gamma_5\sigma_{\mu\nu} C \bar{c}^{T}_n(x) \right] \, , \nonumber\\
J^{PV}_{+,\mu}(x)&=&\frac{\varepsilon^{ijk}\varepsilon^{imn}}{\sqrt{2}}
\Big[s^{T}_j(x)Cc_k(x) \bar{s}_m(x)\gamma_5\gamma_\mu C \bar{c}^{T}_n(x)-s^{T}_j(x)C\gamma_\mu \gamma_5c_k(x)\bar{s}_m(x)C \bar{c}^{T}_n(x) \Big] \, ,\nonumber\\
J^{AA}_{+,\mu\nu}(x)&=&\frac{\varepsilon^{ijk}\varepsilon^{imn}}{\sqrt{2}}
\Big[s^{T}_j(x) C\gamma_\mu c_k(x) \bar{s}_m(x) \gamma_\nu C \bar{c}^{T}_n(x)  +s^{T}_j(x) C\gamma_\nu c_k(x) \bar{s}_m(x) \gamma_\mu C \bar{c}^{T}_n(x) \Big] \, , \nonumber\\
J^{VV}_{+,\mu\nu}(x)&=&\frac{\varepsilon^{ijk}\varepsilon^{imn}}{\sqrt{2}}
\Big[s^{T}_j(x) C\gamma_\mu \gamma_5c_k(x) \bar{s}_m(x) \gamma_5\gamma_\nu C \bar{c}^{T}_n(x)  +s^{T}_j(x) C\gamma_\nu\gamma_5 c_k(x) \bar{s}_m(x) \gamma_5\gamma_\mu C \bar{c}^{T}_n(x) \Big] \, , \nonumber\\
\end{eqnarray}
$\sigma^t_{\mu\nu} =\frac{i}{2}\Big[\gamma^t_\mu, \gamma^t_\nu \Big]$, $\sigma^v_{\mu\nu} =\frac{i}{2}\Big[\gamma^v_\mu, \gamma^t_\nu \Big]$,
$\gamma^v_\mu =  \gamma \cdot t t_\mu$, $\gamma^t_\mu=\gamma_\mu-\gamma \cdot t t_\mu$, $t^\mu=(1,\vec{0})$,  the $i$, $j$, $k$, $m$, $n$ are  color indexes,
    the  charge conjugation matrix $C=i \gamma^2 \gamma^0$, the subscripts $\pm$ stand for the $\pm$ charge conjugation, respectively, the subscripts or superscripts $P$, $S$, $A$($\widetilde{A}$) and $V$($\widetilde{V}$) stand for  the corresponding  diquarks and antidiquarks. For more technical details, one can consult Refs.  \cite{Wang1607-Y4140,WangZG-Di-Y4140,X3915-X4500-EPJA-WZG,X3915-X4500-EPJC-WZG,
    Wang1502-Y4140,WZG-Zcs3985-4110,WZG-HC-spectrum-PRD}.

We would like neglect the superscripts and subscripts, if they are not necessary,   to examine the parity and charge-conjugation of the interpolating currents, then   under parity transform $\widehat{P}$, the quark currents
exhibit  the below  properties,
\begin{eqnarray}\label{P-trans}
\widehat{P} J(x)\widehat{P}^{-1}&=&+J(\tilde{x}) \, , \nonumber\\
\widehat{P} J_\mu(x)\widehat{P}^{-1}&=&-J^\mu(\tilde{x}) \, , \nonumber\\
\widehat{P} J^{S\widetilde{A}}_{\mu\nu}(x)\widehat{P}^{-1}&=&-J_{S\widetilde{A}}^{\mu\nu}(\tilde{x}) \, , \nonumber\\
\widehat{P} J^{AA/VV}_{\mu\nu}(x)\widehat{P}^{-1}&=&+J_{AA/VV}^{\mu\nu}(\tilde{x}) \, ,
\end{eqnarray}
where the coordinates $x^\mu=(t,\vec{x})$ and $\tilde{x}^\mu=(t,-\vec{x})$. While under charge conjugation transform $\widehat{C}$, the quark currents
exhibit the below  properties,
\begin{eqnarray}\label{C-trans}
\widehat{C}J(x)\widehat{C}^{-1}&=&+ J(x) \, , \nonumber\\
\widehat{C}J_{\pm,\mu}(x)\widehat{C}^{-1}&=&\pm J_{\pm,\mu}(x)  \, , \nonumber\\
\widehat{C}J_{\pm,\mu\nu}(x)\widehat{C}^{-1}&=&\pm J_{\pm,\mu\nu}(x)  \, .
\end{eqnarray}
From Eqs.\eqref{P-trans}-\eqref{C-trans}, we can see that all the quark currents have definite parity and charge-conjugation, see Table \ref{Current-Table}.

In this work, we update the old calculations for
  the currents $J_{SS}(x)$, $J_{PP}(x)$ \cite{X3915-X4500-EPJA-WZG}, $J_{AA}(x)$ \cite{X3915-X4500-EPJC-WZG,Wang1502-Y4140}, $J_{VV}(x)$ \cite{X3915-X4500-EPJC-WZG}, $J^{PV}_{+,\mu}(x)$, $J^{PV}_{-,\mu}(x)$, $J^{SA}_{+,\mu}(x)$, $J^{SA}_{-,\mu}(x)$ \cite{Wang1607-Y4140},
$J_{+,\mu}^{\widetilde{V}V}(x)$, $J_{+,\mu}^{\widetilde{A}A}(x)$ \cite{WangZG-Di-Y4140}, $J^{AA}_{+,\mu\nu}(x)$ \cite{Wang1502-Y4140} and $J^{AA}_{-,\mu\nu}(x)$ \cite{WZG-Zcs3985-4110}  to outcome shortcomings in one way  and the other,  and obtain both new analytical and new numerical results for other currents.

\begin{table}
\begin{center}
\begin{tabular}{|c|c|c|c|c|c|c|c|c|}\hline\hline
 $X_c$                      & $J^{PC}$  & Currents       \\ \hline

$[sc]_{S}[\overline{sc}]_{S}$   & $0^{++}$  & $J_{SS}(x)$     \\

$[sc]_{A}[\overline{sc}]_{A}$   & $0^{++}$  & $J_{AA}(x)$               \\

$[sc]_{\tilde{A}}[\overline{sc}]_{\tilde{A}}$  & $0^{++}$  & $J_{\widetilde{A}\widetilde{A}}(x)$   \\

$[sc]_{V}[\overline{sc}]_{V}$   & $0^{++}$  & $J_{VV}(x)$   \\

$[sc]_{\tilde{V}}[\overline{sc}]_{\tilde{V}}$   & $0^{++}$  & $J_{\widetilde{V}\widetilde{V}}(x)$        \\

$[sc]_{P}[\overline{sc}]_{P}$   & $0^{++}$  & $J_{PP}(x)$     \\ \hline

$[sc]_S[\overline{sc}]_{A}-[sc]_{A}[\overline{sc}]_S$  & $1^{+-}$  & $J^{SA}_{-,\mu}(x)$   \\

$[sc]_{A}[\overline{sc}]_{A}$  & $1^{+-}$  & $J^{AA}_{-,\mu\nu}(x)$        \\

$[sc]_S[\overline{sc}]_{\widetilde{A}}-[sc]_{\widetilde{A}}[\overline{sc}]_S$     & $1^{+-}$  & $J^{S\widetilde{A}}_{-,\mu\nu}(x)$     \\

$[sc]_{\widetilde{A}}[\overline{sc}]_{A}-[sc]_{A}[\overline{sc}]_{\widetilde{A}}$ & $1^{+-}$  & $J_{-,\mu}^{\widetilde{A}A}(x)$   \\

$[sc]_{\widetilde{V}}[\overline{sc}]_{V}+[sc]_{V}[\overline{sc}]_{\widetilde{V}}$ & $1^{+-}$  & $J_{-,\mu}^{\widetilde{V}V}(x)$      \\

$[sc]_{V}[\overline{sc}]_{V}$   & $1^{+-}$  & $J^{VV}_{-,\mu\nu}(x)$        \\

$[sc]_P[\overline{sc}]_{V}+[sc]_{V}[\overline{sc}]_P$  & $1^{+-}$  & $J^{PV}_{-,\mu}(x)$    \\ \hline

$[sc]_S[\overline{sc}]_{A}+[sc]_{A}[\overline{sc}]_S$  & $1^{++}$  & $J^{SA}_{+,\mu}(x)$        \\

$[sc]_S[\overline{sc}]_{\widetilde{A}}+[sc]_{\widetilde{A}}[\overline{sc}]_S$     & $1^{++}$  & $J^{S\widetilde{A}}_{+,\mu\nu}(x)$     \\

$[sc]_{\widetilde{V}}[\overline{sc}]_{V}-[sc]_{V}[\overline{sc}]_{\widetilde{V}}$ & $1^{++}$  & $J_{+,\mu}^{\widetilde{V}V}(x)$      \\

$[sc]_{\widetilde{A}}[\overline{sc}]_{A}+[sc]_{A}[\overline{sc}]_{\widetilde{A}}$ & $1^{++}$  & $J_{+,\mu}^{\widetilde{A}A}(x)$       \\

$[sc]_P[\overline{sc}]_{V}-[sc]_{V}[\overline{sc}]_P$   & $1^{++}$  & $J^{PV}_{+,\mu}(x)$      \\ \hline

$[sc]_{A}[\overline{sc}]_{A}$  & $2^{++}$  & $J^{AA}_{+,\mu\nu}(x)$       \\

$[sc]_{V}[\overline{sc}]_{V}$  & $2^{++}$  & $J^{VV}_{+,\mu\nu}(x)$        \\
\hline\hline
\end{tabular}
\end{center}
\caption{ Correspondence between the diquark structures and currents. }\label{Current-Table}
\end{table}

At the hadron side, we adopt the standard procedure \cite{SVZ79,Reinders85}, and insert  a complete set of intermediate hadronic states with the same quantum numbers as the quark currents $J(x)$, $J_\mu(x)$ and $J_{\mu\nu}(x)$ into the
correlation functions $\Pi(p)$, $\Pi_{\mu\nu}(p)$ and $\Pi_{\mu\nu\alpha\beta}(p)$   according to the quark-hadron duality to get the hadronic spectral
representation, and isolate the ground state hidden-charm-hidden-strange tetraquark contributions,
\begin{eqnarray}
\Pi(p)&=&\frac{\lambda_{X^+}^2}{M_{X^+}^2-p^2} +\cdots \nonumber\\
&=&\Pi_{+}(p^2) \, ,\nonumber\\
\Pi_{\mu\nu}(p)&=&\frac{\lambda_{X^+}^2}{M_{X^+}^2-p^2}\left( -g_{\mu\nu}+\frac{p_{\mu}p_{\nu}}{p^2}\right) +\cdots \nonumber\\
&=&\Pi_{+}(p^2)\left( -g_{\mu\nu}+\frac{p_{\mu}p_{\nu}}{p^2}\right)+\cdots \, ,\nonumber
\end{eqnarray}
\begin{eqnarray}
\Pi^{AA,-}_{\mu\nu\alpha\beta}(p)&=&\frac{\lambda_{ X^+}^2}{M_{X^+}^2\left(M_{X^+}^2-p^2\right)}\left(p^2g_{\mu\alpha}g_{\nu\beta} -p^2g_{\mu\beta}g_{\nu\alpha} -g_{\mu\alpha}p_{\nu}p_{\beta}-g_{\nu\beta}p_{\mu}p_{\alpha}+g_{\mu\beta}p_{\nu}p_{\alpha}+g_{\nu\alpha}p_{\mu}p_{\beta}\right) \nonumber\\
&&+\frac{\lambda_{ X^-}^2}{M_{X^-}^2\left(M_{X^-}^2-p^2\right)}\left( -g_{\mu\alpha}p_{\nu}p_{\beta}-g_{\nu\beta}p_{\mu}p_{\alpha}+g_{\mu\beta}p_{\nu}p_{\alpha}+g_{\nu\alpha}p_{\mu}p_{\beta}\right) +\cdots  \nonumber\\
&=&\widetilde{\Pi}_{+}(p^2)\left(p^2g_{\mu\alpha}g_{\nu\beta} -p^2g_{\mu\beta}g_{\nu\alpha} -g_{\mu\alpha}p_{\nu}p_{\beta}-g_{\nu\beta}p_{\mu}p_{\alpha}+g_{\mu\beta}p_{\nu}p_{\alpha}+g_{\nu\alpha}p_{\mu}p_{\beta}\right) \nonumber\\
&&+\widetilde{\Pi}_{-}(p^2)\left( -g_{\mu\alpha}p_{\nu}p_{\beta}-g_{\nu\beta}p_{\mu}p_{\alpha}+g_{\mu\beta}p_{\nu}p_{\alpha}+g_{\nu\alpha}p_{\mu}p_{\beta}\right) \, ,\nonumber\\
\Pi^{S\widetilde{A},\pm}_{\mu\nu\alpha\beta}(p)&=&\frac{\lambda_{ X^-}^2}{M_{X^-}^2\left(M_{X^-}^2-p^2\right)}\left(p^2g_{\mu\alpha}g_{\nu\beta} -p^2g_{\mu\beta}g_{\nu\alpha} -g_{\mu\alpha}p_{\nu}p_{\beta}-g_{\nu\beta}p_{\mu}p_{\alpha}+g_{\mu\beta}p_{\nu}p_{\alpha}+g_{\nu\alpha}p_{\mu}p_{\beta}\right) \nonumber\\
&&+\frac{\lambda_{ X^+}^2}{M_{X^+}^2\left(M_{X^+}^2-p^2\right)}\left( -g_{\mu\alpha}p_{\nu}p_{\beta}-g_{\nu\beta}p_{\mu}p_{\alpha}+g_{\mu\beta}p_{\nu}p_{\alpha}+g_{\nu\alpha}p_{\mu}p_{\beta}\right) +\cdots  \nonumber\\
&=&\widetilde{\Pi}_{-}(p^2)\left(p^2g_{\mu\alpha}g_{\nu\beta} -p^2g_{\mu\beta}g_{\nu\alpha} -g_{\mu\alpha}p_{\nu}p_{\beta}-g_{\nu\beta}p_{\mu}p_{\alpha}+g_{\mu\beta}p_{\nu}p_{\alpha}+g_{\nu\alpha}p_{\mu}p_{\beta}\right) \nonumber\\
&&+\widetilde{\Pi}_{+}(p^2)\left( -g_{\mu\alpha}p_{\nu}p_{\beta}-g_{\nu\beta}p_{\mu}p_{\alpha}+g_{\mu\beta}p_{\nu}p_{\alpha}+g_{\nu\alpha}p_{\mu}p_{\beta}\right) \, , \nonumber\\
\Pi_{\mu\nu\alpha\beta}^{AA,+}(p)&=&\frac{\lambda_{ X^+}^2}{M_{X^+}^2-p^2}\left( \frac{\widetilde{g}_{\mu\alpha}\widetilde{g}_{\nu\beta}+\widetilde{g}_{\mu\beta}\widetilde{g}_{\nu\alpha}}{2}-\frac{\widetilde{g}_{\mu\nu}\widetilde{g}_{\alpha\beta}}{3}\right) +\cdots \, \, , \nonumber \\
&=&\Pi_{+}(p^2)\left( \frac{\widetilde{g}_{\mu\alpha}\widetilde{g}_{\nu\beta}+\widetilde{g}_{\mu\beta}\widetilde{g}_{\nu\alpha}}{2}-\frac{\widetilde{g}_{\mu\nu}\widetilde{g}_{\alpha\beta}}{3}\right) +\cdots\, ,
\end{eqnarray}
where $\widetilde{g}_{\mu\nu}=g_{\mu\nu}-\frac{p_{\mu}p_{\nu}}{p^2}$.  We introduce  the superscripts $\pm$ in the correlation functions $\Pi^{AA,-}_{\mu\nu\alpha\beta}(p)$, $\Pi^{S\widetilde{A},\pm}_{\mu\nu\alpha\beta}(p)$ and $\Pi_{\mu\nu\alpha\beta}^{AA,+}(p)$ to indicate the $\pm$ charge conjugation, respectively,
  introduce the superscripts (subscripts) $\pm$ in the  tetraquark states $X^{\pm}$ (correlation functions $\Pi_{\pm}(p^2)$, $\widetilde{\Pi}_{\pm}(p^2)$) to indicate  the $\pm$ parity, respectively. The correlation functions $\Pi^{VV,\pm}_{\mu\nu\alpha\beta}(p)$ and $\Pi^{AA,\pm}_{\mu\nu\alpha\beta}(p)$ have the same tensor structures, and are analyzed in the same way.  We define the pole residues  $\lambda_{X^\pm}$ by conventional rules,
\begin{eqnarray}
 \langle 0|J(0)|X^+(p)\rangle &=&\lambda_{X^+}\, , \nonumber\\
 \langle 0|J_\mu(0)|X^+(p)\rangle &=&\lambda_{X^+}\varepsilon_\mu\, , \nonumber\\
  \langle 0|J_{\pm,\mu\nu}^{S\widetilde{A}}(0)|X^-(p)\rangle &=& \frac{\lambda_{X^-}}{M_{X^-}} \, \varepsilon_{\mu\nu\alpha\beta} \, \varepsilon^{\alpha}p^{\beta}\, , \nonumber\\
 \langle 0|J_{\pm,\mu\nu}^{S\widetilde{A}}(0)|X^+(p)\rangle &=&\frac{\lambda_{X^+}}{M_{X^+}} \left(\varepsilon_{\mu}p_{\nu}-\varepsilon_{\nu}p_{\mu} \right)\, , \nonumber\\
  \langle 0|J_{-,\mu\nu}^{AA/VV}(0)|X^+(p)\rangle &=& \frac{\lambda_{X^+}}{M_{X^+}} \, \varepsilon_{\mu\nu\alpha\beta} \, \varepsilon^{\alpha}p^{\beta}\, , \nonumber\\
 \langle 0|J_{-,\mu\nu}^{AA/VV}(0)|X^-(p)\rangle &=&\frac{\lambda_{X^-}}{M_{X^-}} \left(\varepsilon_{\mu}p_{\nu}-\varepsilon_{\nu}p_{\mu} \right)\, , \nonumber\\
  \langle 0|J_{+,\mu\nu}^{AA/VV}(0)|X^+(p)\rangle &=& \lambda_{X^+}\, \varepsilon_{\mu\nu} \, ,
\end{eqnarray}
where the  $\varepsilon_{\mu/\alpha}$ and $\varepsilon_{\mu\nu}$ are the corresponding polarization vectors. We choose the components $\Pi_{+}(p^2)$ and $p^2\widetilde{\Pi}_{+}(p^2)$ to explore the  tetraquark states with the quantum numbers $J^{PC}=0^{++}$, $1^{+-}$, $1^{++}$ and $2^{++}$ without contaminations from  other states, respectively.

We always perform Fierz transformation routinely to transform  the diquark-antidiquark type four-quark currents  into a particular  superposition of the color-singlet-color-singlet type  currents \cite{Wang-tetra-formula}, for example,
\begin{eqnarray}\label{Fierz}
2\sqrt{2}J^{SA}_{+,\mu}&=&\,i\bar{c}i\gamma_5 c\,\bar{s}\gamma_\mu s-i\bar{c} \gamma_\mu c\,\bar{s}i\gamma_5 s+\bar{c} s\,\bar{s}\gamma_\mu\gamma_5 c
-\bar{c} \gamma_\mu \gamma_5s\,\bar{s}c  \nonumber\\
&&  - i\bar{c}\gamma^\nu\gamma_5c\, \bar{s}\sigma_{\mu\nu}s+i\bar{c}\sigma_{\mu\nu}c\, \bar{s}\gamma^\nu\gamma_5s
- i \bar{c}\sigma_{\mu\nu}\gamma_5s\,\bar{s}\gamma^\nu c+i\bar{c}\gamma^\nu s\, \bar{s}\sigma_{\mu\nu}\gamma_5c     \, .
\end{eqnarray}
In Refs.\cite{WangZG-Landau,Wang-Two-particle}, we provide  detailed discussions to illustrate that  the two-meson scattering states play a minor important role and cannot saturate the QCD sum rules alone (or by themselves),  while the tetraquark (molecular) states play an un-substitutable role, we can saturate the QCD sum rules with or without the two-particle scattering state contributions, it is feasible and reasonable to apply the QCD sum rules to investigate the  tetraquark (molecular) states.

Now we revisit the QCD side and
make great efforts to get the QCD spectral representation. We accomplish the operator product expansion  up to the vacuum condensates of dimension $10$ consistently according to our unique counting/truncation rules \cite{WZG-tetra-psedo-NPB,WZG-HC-spectrum-PRD,WZG-EPJC-P-wave,WZG-EPJC-P-2P}, and take into account the vacuum condensates $\langle\bar{s}s\rangle$, $\langle\frac{\alpha_{s}GG}{\pi}\rangle$, $\langle\bar{s}g_{s}\sigma Gs\rangle$, $\langle\bar{s}s\rangle^2$, $g_s^2\langle\bar{s}s\rangle^2$,
$\langle\bar{s}s\rangle \langle\frac{\alpha_{s}GG}{\pi}\rangle$,  $\langle\bar{s}s\rangle  \langle\bar{s}g_{s}\sigma Gs\rangle$,
$\langle\bar{s}g_{s}\sigma Gs\rangle^2$ and $\langle\bar{s}s\rangle^2 \langle\frac{\alpha_{s}GG}{\pi}\rangle$ because  we take  the truncation ${\mathcal{O}}(\alpha_s^k)$ with $k\leq 1$ for the quark-gluon operators, just like what have been done in our previous works therefore we have uniform standard to judge the reliability of our calculations,  then we obtain the QCD spectral densities $\rho_{QCD}(s)$ through dispersion relation straightforwardly. We  recalculate the higher  dimensional vacuum condensates by applying the identity  $t^a_{ij}t^a_{mn}=-\frac{1}{6}\delta_{ij}\delta_{mn}+\frac{1}{2}\delta_{jm}
\delta_{in}$ in the color space rigorously, and obtain slightly  different analytical expressions  compared with the old ones \cite{Wang1607-Y4140,X3915-X4500-EPJA-WZG,X3915-X4500-EPJC-WZG,Wang1502-Y4140}.
In calculations, we take the full quark propagators,
\begin{eqnarray}
S^{ij}(x)&=& \frac{i\delta_{ij}\!\not\!{x}}{ 2\pi^2x^4}
-\frac{\delta_{ij}m_s}{4\pi^2x^2}-\frac{\delta_{ij}\langle
\bar{s}s\rangle}{12} +\frac{i\delta_{ij}\!\not\!{x}m_s
\langle\bar{s}s\rangle}{48}-\frac{\delta_{ij}x^2\langle \bar{s}g_s\sigma Gs\rangle}{192}+\frac{i\delta_{ij}x^2\!\not\!{x} m_s\langle \bar{s}g_s\sigma
 Gs\rangle }{1152}\nonumber\\
&& -\frac{ig_s G^{a}_{\alpha\beta}t^a_{ij}(\!\not\!{x}
\sigma^{\alpha\beta}+\sigma^{\alpha\beta} \!\not\!{x})}{32\pi^2x^2} -\frac{i\delta_{ij}x^2\!\not\!{x}g_s^2\langle \bar{s} s\rangle^2}{7776} -\frac{\delta_{ij}x^4\langle \bar{s}s \rangle\langle g_s^2 GG\rangle}{27648}-\frac{1}{8}\langle\bar{s}_j\sigma^{\mu\nu}s_i \rangle \sigma_{\mu\nu} \nonumber\\
&&   -\frac{1}{4}\langle\bar{s}_j\gamma^{\mu}s_i\rangle \gamma_{\mu }+\cdots \, ,
\end{eqnarray}
\begin{eqnarray}
S_Q^{ij}(x)&=&\frac{i}{(2\pi)^4}\int d^4k e^{-ik \cdot x} \left\{
\frac{\delta_{ij}}{\!\not\!{k}-m_Q}
-\frac{g_sG^n_{\alpha\beta}t^n_{ij}}{4}\frac{\sigma^{\alpha\beta}(\!\not\!{k}+m_Q)+(\!\not\!{k}+m_Q)
\sigma^{\alpha\beta}}{(k^2-m_Q^2)^2}\right.\nonumber\\
&&\left. +\frac{g_s D_\alpha G^n_{\beta\lambda}t^n_{ij}(f^{\lambda\beta\alpha}+f^{\lambda\alpha\beta}) }{3(k^2-m_Q^2)^4}-\frac{g_s^2 (t^at^b)_{ij} G^a_{\alpha\beta}G^b_{\mu\nu}(f^{\alpha\beta\mu\nu}+f^{\alpha\mu\beta\nu}+f^{\alpha\mu\nu\beta}) }{4(k^2-m_Q^2)^5}+\cdots\right\} \, ,\nonumber\\
f^{\lambda\alpha\beta}&=&(\!\not\!{k}+m_Q)\gamma^\lambda(\!\not\!{k}+m_Q)\gamma^\alpha(\!\not\!{k}+m_Q)\gamma^\beta(\!\not\!{k}+m_Q)\, ,\nonumber\\
f^{\alpha\beta\mu\nu}&=&(\!\not\!{k}+m_Q)\gamma^\alpha(\!\not\!{k}+m_Q)\gamma^\beta(\!\not\!{k}+m_Q)\gamma^\mu(\!\not\!{k}+m_Q)\gamma^\nu(\!\not\!{k}+m_Q)\, ,
\end{eqnarray}
and   $D_\alpha=\partial_\alpha-ig_sG^n_\alpha t^n$ \cite{WangHuangTao-3900,Reinders85,Pascual-1984}.

We introduce the new terms $\langle\bar{s}_j\sigma_{\mu\nu}s_i \rangle$ and $\langle\bar{s}_j\gamma_{\mu}s_i\rangle$ in the full light-quark propagators by applying the Fierz transformation \cite{WangHuangTao-3900}, which  absorb the gluons  emitted from the other (irrespective of heavy or light) quark lines and contribute to the mixed condensate and four-quark condensate $\langle\bar{s}g_s\sigma G s\rangle$ and $g_s^2\langle\bar{s}s\rangle^2$, respectively.
 The typical vacuum condensate $g_s^2\langle \bar{s}s\rangle^2$ originates from the vacuum matrix elements
$\langle \bar{s}\gamma_\mu t^a s g_s D_\eta G^a_{\lambda\tau}\rangle$, $\langle\bar{s}_jD^{\dagger}_{\mu}D^{\dagger}_{\nu}D^{\dagger}_{\alpha}s_i\rangle$  and
$\langle\bar{s}_jD_{\mu}D_{\nu}D_{\alpha}s_i\rangle$  rather than originates from the radiative  $\mathcal{O}(\alpha_s)$ corrections for the vacuum  condensate $\langle \bar{s}s\rangle^2$. The strong fine-structure  constant $\alpha_s(\mu)=\frac{g_s^2(\mu)}{4\pi}$ appears  at the tree level, which is energy scale dependent and warrants the necessity and
legitimacy of applying the (modified) energy scale formula.
The contributions of the typical condensate  $g_s^2\langle\bar{s}s\rangle^2$ are tiny and neglected in most QCD sum rules, because  they are not companied with the Borel parameters of the forms $\frac{1}{T^2}$, $\frac{1}{T^4}$, $\frac{1}{T^6}$, $\cdots$  to magnify  themselves at  small $T^2$ therefore affect choosing the Borel windows.

According to the quark-hadron duality, the basic assumption of the QCD sum rules, we match  the hadron side with the QCD  side for the components $\Pi_{+}(p^2)$ and $p^2\widetilde{\Pi}_{+}(p^2)$ in the spectral representation below the continuum thresholds   $s_0$ and accomplish the Borel transform  in regard to
the $P^2=-p^2$ to get   the main  QCD sum rules:
\begin{eqnarray}\label{QCDSR}
\lambda^2_{X^+}\, \exp\left(-\frac{M^2_{X^+}}{T^2}\right)= \int_{4m_c^2}^{s_0} ds\, \rho_{QCD}(s) \, \exp\left(-\frac{s}{T^2}\right) \, .
\end{eqnarray}

As the last step analytical calculations, we differentiate the main QCD sum rules in Eq.\eqref{QCDSR} with respect to  the variable $\tau=\frac{1}{T^2}$,  and acquire the QCD sum rules for  the masses of the $cs\overline{cs}$  tetraquark states $X^+$ with the positive parity,
 \begin{eqnarray}
 M^2_{X^+}&=& -\frac{\int_{4m_c^2}^{s_0} ds\frac{d}{d \tau}\rho_{QCD}(s)\exp\left(-\tau s \right)}{\int_{4m_c^2}^{s_0} ds \rho_{QCD}(s)\exp\left(-\tau s\right)}\, .
\end{eqnarray}

\section{Numerical results and discussions}
 It is well known that the  $\overline{MS}$ quark masses and the vacuum condensates depend on the energy scale $\mu$, accordingly,  the QCD spectral densities $\rho(s,\mu)$, the thresholds $4m_c^2(\mu)$, and the integral intervals between the thresholds $4m_c^2(\mu)$ and continuum thresholds $s_0$, are all  dependent on the energy scale $\mu$. In numerical calculations, we observe that the predictions are sensitive to the energy scale, small variations of the energy scale can lead to rather different results, and weaken the reliability of the predictions. In a short summary,    we cannot obtain energy scale independent QCD sum rules, and have to choose the suitable or pertinent   energy scales to extract robust  tetraquark masses and pole residues.

Let us write down the energy-scale dependence of  the input parameters,
\begin{eqnarray}
\langle\bar{s}s \rangle(\mu)&=&\langle\bar{s}s \rangle({\rm 1GeV})\left[\frac{\alpha_{s}({\rm 1GeV})}{\alpha_{s}(\mu)}\right]^{\frac{12}{33-2n_f}}\, , \nonumber\\
 \langle\bar{s}g_s \sigma Gs \rangle(\mu)&=&\langle\bar{s}g_s \sigma Gs \rangle({\rm 1GeV})\left[\frac{\alpha_{s}({\rm 1GeV})}{\alpha_{s}(\mu)}\right]^{\frac{2}{33-2n_f}}\, , \nonumber\\
 m_c(\mu)&=&m_c(m_c)\left[\frac{\alpha_{s}(\mu)}{\alpha_{s}(m_c)}\right]^{\frac{12}{33-2n_f}} \, ,\nonumber\\
m_s(\mu)&=&m_s({\rm 2GeV} )\left[\frac{\alpha_{s}(\mu)}{\alpha_{s}({\rm 2GeV})}\right]^{\frac{12}{33-2n_f}}\, ,\nonumber\\
\alpha_s(\mu)&=&\frac{1}{b_0t}\left[1-\frac{b_1}{b_0^2}\frac{\log t}{t} +\frac{b_1^2(\log^2{t}-\log{t}-1)+b_0b_2}{b_0^4t^2}\right]\, ,
\end{eqnarray}
 where $t=\log \frac{\mu^2}{\Lambda_{QCD}^2}$, $b_0=\frac{33-2n_f}{12\pi}$, $b_1=\frac{153-19n_f}{24\pi^2}$, $b_2=\frac{2857-\frac{5033}{9}n_f+\frac{325}{27}n_f^2}{128\pi^3}$,  $\Lambda_{QCD}=210\,\rm{MeV}$, $292\,\rm{MeV}$  and  $332\,\rm{MeV}$ for the flavors  $n_f=5$, $4$ and $3$, respectively  \cite{PDG,Narison-mix}. Because  the $c$-quark is concerned, we choose the flavor number to be $n_f=4$.

 At the beginning  points, we adopt  the conventional/commonly-used values  $\langle
\bar{q}q \rangle=-(0.24\pm 0.01\, \rm{GeV})^3$, $\langle
\bar{s} s \rangle=(0.8 \pm 0.1)\langle \bar{q}q \rangle$,  $\langle
\bar{s}g_s\sigma G s \rangle=m_0^2\langle \bar{s}s \rangle$,
$m_0^2=(0.8 \pm 0.1)\,\rm{GeV}^2$,  $\langle \frac{\alpha_s
GG}{\pi}\rangle=(0.012\pm0.004)\,\rm{GeV}^4 $    at the typical energy scale  $\mu=1\, \rm{GeV}$
\cite{SVZ79,Reinders85,Colangelo-Review}, and adopt  the $\overline{MS}$
(modified-minimal-subtraction) masses   $m_{c}(m_c)=(1.275\pm0.025)\,\rm{GeV}$ and $m_{s}({\rm 2 GeV})=(0.095\pm0.005)\,\rm{GeV}$ from the Particle Data Group \cite{PDG}.

In the work on hand, we apply the modified energy scale formula $\mu=\sqrt{M^2_{X/Y/Z}-(2{\mathbb{M}}_c)^2}-2{\mathbb{M}}_s$ to pick out the suitable/reasonable  energy scales of the QCD spectral densities \cite{Wang-tetra-formula,WZG-mole-IJMPA,WZG-XQ-mole-penta,WZG-XQ-mole-EPJA,
WZG-tetra-psedo-NPB,WZG-Zcs3985-4110,WZG-NPB-cscs}, where the ${\mathbb{M}}_c$ and ${\mathbb{M}}_s$ are the effective $c$ and $s$-quark masses respectively, and have universal values to be commonly used elsewhere. We adopt  the updated value  ${\mathbb{M}}_c=1.82\,\rm{GeV}$ \cite{WangEPJC-1601}, and take  the collective (or net)  light-flavor $SU(3)$-breaking effects into account  by
employing an  effective $s$-quark mass ${\mathbb{M}}_s=0.20\,\rm{GeV}$ ($0.12\,\rm{GeV}$) for the S-wave (P-wave) tetraquark  states  \cite{WZG-mole-IJMPA,WZG-XQ-mole-penta,WZG-XQ-mole-EPJA,
WZG-tetra-psedo-NPB,WZG-Zcs3985-4110,WZG-NPB-cscs},  as a matter of fact, we can adopt  the value  ${\mathbb{M}}_s=0.20\,\rm{GeV}$ uniformly.
 The (modified) energy scale formula  can  enhance  the ground state contributions significantly  and ameliorate  the convergent behaviors of the operator product expansion significantly, therefore,  it is more easy and more reliable   to obtain  the lowest (ground state) tetraquark  (molecule) masses.

In the scenario of diquark-antidiquark type tetraquark states, we can tentatively identify  the $X(3915)$ and $X(4500)$ as the 1S and 2S hidden-charm-hidden-strange tetraquark  states with the quantum numbers $J^{PC}=0^{++}$ \cite{X4140-tetraquark-Lebed,X3915-X4500-EPJC-WZG}, identify
the $Z_c(3900)$ and $Z_c(4430)$   as  the 1S and 2S  hidden-charm tetraquark states with the quantum numbers $J^{PC}=1^{+-}$, respectively according to the
analogous decays $Z_c^\pm(3900)\to J/\psi\pi^\pm$, $Z_c^\pm(4430)\to\psi^\prime\pi^\pm$ \cite{Maiani-II-type,Nielsen-1401,WangZG-Z4430-CTP},   identify the $Z_c(4020)$ and $Z_c(4600)$ as the 1S and 2S hidden-charm tetraquark states with the quantum numbers $J^{PC}=1^{+-}$, respectively  \cite{ChenHX-Z4600-A,WangZG-axial-Z4600}, and identify the $X(4140)$ and $X(4685)$ as the 1S and 2S hidden-charm-hidden-strange  tetraquark states with the quantum numbers  $J^{PC}=1^{++}$, respectively \cite{WZG-X4140-X4685}. We draw the conclusion approximately that the mass gaps between the 1S and 2S tetraquark states  are about $0.57\sim 0.59 \,\rm{GeV}$.

In 2022,  the LHCb collaboration observed the $X(3960)$ in the $D_s^+D_s^-$ invariant mass spectrum with the favored identification $J^{PC}=0^{++}$  \cite{LHCb3960-2022}, it is more reasonable to identify the $X(3960)$ (instead of the $X(3915)$) as the ground state $cs\bar{c}\bar{s}$ tetraquark state, then the mass gap $M_{X(4500)}-M_{X(3960)}=550\,\rm{MeV}$ \cite{LHCb-16061,LHCb-16062}. Furthermore,  the BESIII collaboration studied the processes  $e^+e^-\to\omega X(3872)$ and $\gamma X(3872)$,  and observed that the relatively large cross section for the $e^+e^-\to\omega X(3872)$ process is mainly attributed to the  enhancement around 4.75 GeV, which maybe  indicate a potential structure in the $e^+e^-\to\omega X(3872)$  cross section  \cite{Y4750-BESIII}. In Ref.\cite{WZG-EPJC-P-2P},   we choose the diquark-antidiquark type four-quark currents having  an explicit P-wave between the diquark and antidiquark pairs to diagnose the 1P and 2P hidden-charm tetraquark states with the quantum numbers $J^{PC}=1^{--}$. And we observe that there really  exists a hidden-charm  tetraquark   state with the quantum numbers $J^{PC}=1^{--}$ at the energy about $4.75\,\rm{GeV}$ as the 2P state to account for the BESIII data, the mass gap $M_{Y(4750)}-M_{Y(4220/4260)}=510\,\rm{MeV}$. In a short summary, the mass gaps between the 1S and 2S (or 1P and 2P) could also be $0.50\sim 0.55\,\rm{GeV}$.

In the present work, we would like provide  two sets of data for the scalar tetraquark states, one set data is based on the continuum threshold parameters about  $\sqrt{s_0}=M_X+0.60\pm 0.10\,\rm{GeV}$ with the central values $\sqrt{s_0}<M_Y+0.60\,\rm{GeV}$ (the criterion is also adopted for the tetraquark states with the $J^{PC}=1^{+-}$, $1^{++}$ and $2^{++}$), the other set data is based on  the  continuum threshold parameters about $\sqrt{s_0}=M_X+0.55\pm 0.10\,\rm{GeV}$ with the central values $\sqrt{s_0}<M_Y+0.55\,\rm{GeV}$, which will be characterized by the additional symbol $*$ in Tables \ref{BorelP-ss}--\ref{Identifications-Table-ss}.

At this step,  we change  the continuum threshold parameters $s_0$ and Borel parameters $T^2$ to meet with the following four   criteria:\\
$\bullet$ Pole (in other words, ground state) dominance at the hadron  side;\\
$\bullet$ Convergence of the operator product expansion at the QCD side;\\
$\bullet$ Appearance of the enough flat Borel platforms;\\
$\bullet$ Satisfying the  modified energy scale formula,\\
  via trial  and error.
We define the pole contributions (PC)  routinely,
\begin{eqnarray}
{\rm{PC}}&=&\frac{\int_{4m_{c}^{2}}^{s_{0}}ds\rho_{QCD}\left(s\right)\exp\left(-\frac{s}{T^{2}}\right)} {\int_{4m_{c}^{2}}^{\infty}ds\rho_{QCD}\left(s\right)\exp\left(-\frac{s}{T^{2}}\right)}\, ,
\end{eqnarray}
 and define the contributions of the vacuum condensates $D(n)$ of dimension $n$ routinely,
\begin{eqnarray}
D(n)&=&\frac{\int_{4m_{c}^{2}}^{s_{0}}ds\rho_{QCD,n}(s)\exp\left(-\frac{s}{T^{2}}\right)}
{\int_{4m_{c}^{2}}^{s_{0}}ds\rho_{QCD}\left(s\right)\exp\left(-\frac{s}{T^{2}}\right)}\, ,
\end{eqnarray}
because  we merely intend to analyze the 1S tetraquark contributions.
The pole dominance and convergence of the operator product expansion  are two elementary  requirements, we should fulfill
those  constraints strictly   to get robust  QCD sum rules.

At the QCD side of the $\Pi(p)$, $\Pi_{\mu\nu}(p)$ and $\Pi_{\mu\nu\alpha\beta}(p)$, there are  two heavy  quark propagators and two light quark propagators after accomplishing the Wick's contractions.  Let us consider the typical case,   each heavy quark line
radiates a gluon and each light quark line provides  a quark-antiquark  pair, then there is a quark-gluon  operator $GG\bar{s}s \bar{s}s$   having  dimension 10, which leads to the vacuum condensates  $\langle\bar{s}g_{s}\sigma Gs\rangle^2$ and $\langle\bar{s}s\rangle^2 \langle\frac{\alpha_{s}GG}{\pi}\rangle$ under the vacuum saturation assumption, thus we should take account of  the vacuum condensates up to dimension 10 to estimate  the convergent behavior
believably. The vacuum condensates  $\langle\bar{s}g_{s}\sigma Gs\rangle^2$ and $\langle\bar{s}s\rangle^2 \langle\frac{\alpha_{s}GG}{\pi}\rangle$ are companied  with the negative power of the Borel parameter $\frac{1}{T^2}$, $\frac{1}{T^4}$,  $\frac{1}{T^6}$  and  $\frac{1}{T^8}$, they play a crucial important role in choosing the Borel windows, and   we require  the enough small contributions $|D(10)|\sim 1\%$ in the Borel windows, just like in Refs.\cite{WZG-mole-IJMPA,WZG-XQ-mole-EPJA,
WZG-tetra-psedo-NPB,WZG-HC-spectrum-PRD}.

At last, we obtain the Borel windows, continuum threshold parameters, energy scales of the QCD spectral densities,  pole contributions, and  contributions of the vacuum condensates of dimension $10$, and  show them clearly  in Table \ref{BorelP-ss}.
From the Table,  we can see
evidently that the pole (or ground state) contributions are about $(40-60)\%$, the largest pole contributions for the multiquark states  reached   up to today, the pole dominance requirement  can be certainly  fulfilled.
The contributions of the vacuum condensates of dimension $10$ (the highest dimension) are $|D(10)|\ll 1\%$ except for $|D(10)|\leq 1 \%$ for the $[sc]_{A}[\overline{sc}]_{A}{}^*$ tetraquark state with the quantum numbers  $J^{PC}=1^{+-}$, the convergent behaviors of the operator product  expansion are all excellent.

\begin{figure}
 \centering
 \includegraphics[totalheight=7cm,width=10cm]{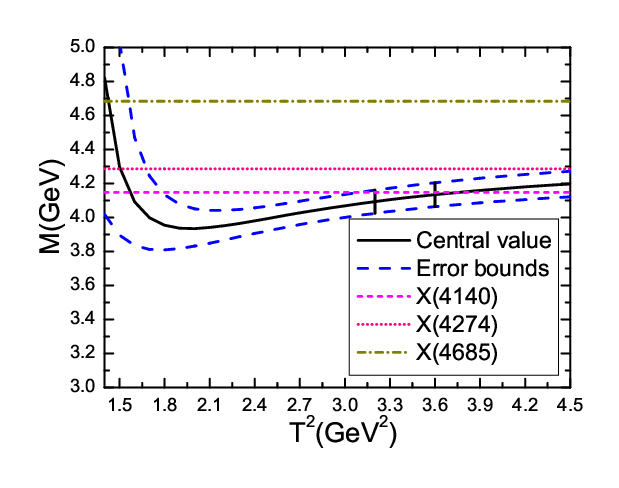}
 \caption{ The mass of the  $[sc]_S[\overline{sc}]_{A}+[sc]_{A}[\overline{sc}]_S$ tetraquark state   with variations  of the Borel parameter $T^2$.  }\label{mass-1-fig}
\end{figure}

We take into account all the uncertainties of the relevant  parameters and acquire  the masses and pole residues of the  hidden-charm-hidden-strange  tetraquark states with the quantum numbers $J^{PC}=0^{++}$, $1^{+-}$, $1^{++}$ and $2^{++}$, and present them  explicitly in Table \ref{mass-Table-ss}. From  Tables \ref{BorelP-ss}--\ref{mass-Table-ss}, we can observe evidently (through extremely simple computation) that the modified energy scale formula $\mu=\sqrt{M^2_{X/Y/Z}-(2{\mathbb{M}}_c)^2}-2{\mathbb{M}}_s$ is obeyed  satisfactorily.
 In  Fig.\ref{mass-1-fig}, we plot the mass of the  $[sc]_S[\overline{sc}]_{A}+[sc]_{A}[\overline{sc}]_S$ tetraquark state  having the quantum numbers $J^{PC}=1^{++}$ with variations of the Borel parameter at much larger range than the Borel widow as an example, where we also give the experimental values of the masses of the $X(4140)$, $X(4274)$ and $X(4685)$ for comparison. There really appear perfectly
smooth  platform in the Borel window, where the two short perpendicular lines
mark the boundaries of the Borel window.  Therefore, we draw the conclusion confidently that the four elementary criteria/requirements are all met with, and  expect to reach  reliable/robust predictions.

In Table \ref{Identifications-Table-ss}, we provide the probable  identifications of the 1S (also 2S) hidden-charm-hidden-strange tetraquark states with the existing $X$ states.
There exist   one  tetraquark candidate with the quantum numbers $J^{PC}=0^{++}$ for the $X(3960)$ and $X(4500)$,
one  tetraquark candidate with the  quantum numbers $J^{PC}=1^{++}$ for the $X(4274)$,
two  tetraquark candidates with the  quantum numbers $J^{PC}=1^{++}$ for the $X(4140)$ and $X(4685)$,
three tetraquark candidates with the  quantum numbers $J^{PC}=0^{++}$ for the $X(4700)$.

The prediction $3.99\pm0.09\,\rm{GeV}$ for the mass of the $[sc]_{S}[\overline{sc}]_{S}{}^*$ tetraquark state with the quantum numbers $J^{PC}=0^{++}$  favors identifying the $X(3960)$ and $X(4500)$  as the 1S and 2S tetraquark states respectively with the energy gap about $0.55\,\rm{GeV}$ from the LHCb collaboration \cite{LHCb-16061,LHCb-16062,LHCb3960-2022} or $0.52\,\rm{GeV}$ from the Particle Data Group \cite{PDG}. In our previous computation  \cite{X3915-X4500-EPJC-WZG}, the predictions
$M_X=3.92^{+0.19}_{-0.18}\,\rm{GeV}$ and $4.50^{+0.08}_{-0.09}\,\rm{GeV}$ favor identifying the $X(3915)$ and $X(4500)$ as the 1S and 2S $[sc]_{A}[\overline{sc}]_{A}$ tetraquark states with the quantum numbers $J^{PC}=0^{++}$, now such identifications are  superseded.

The 1S $[sc]_{A}[\overline{sc}]_{A}$ and $[sc]_{\tilde{A}}[\overline{sc}]_{\tilde{A}}$ tetraquark  states with the  quantum numbers $J^{PC}=0^{++}$  have the masses $4.13\pm0.09\,\rm{GeV}$    and $4.16\pm0.09\,\rm{GeV}$, respectively, which favors identifying the  $X(4700)$ as the first radial excitation of the $[sc]_{A}[\overline{sc}]_{A}$ or $[sc]_{\tilde{A}}[\overline{sc}]_{\tilde{A}}$ tetraquark state approximately. On the other hand, we get the prediction $M_X=4.70\pm0.09\,\rm{GeV}$ for the $[sc]_{V}[\overline{sc}]_{V}{}^*$ tetraquark state  with the quantum numbers  $J^{PC}=0^{++}$, which is in excellent agreement with our previous computation   $M_X=4.70^{+0.08}_{-0.09}\,\rm{GeV}$ \cite{X3915-X4500-EPJC-WZG}, and favors identifying the $X(4700)$ as the $[sc]_{V}[\overline{sc}]_{V}{}^*$ tetraquark state with the quantum numbers $J^{PC}=0^{++}$, this probable identification still survives.

The predictions $M_X=4.11\pm0.09\,\rm{GeV}$ and $4.17\pm0.09\,\rm{GeV}$ for the 1S  $[sc]_S[\overline{sc}]_{A}+[sc]_{A}[\overline{sc}]_S$  and
$[sc]_S[\overline{sc}]_{\widetilde{A}}+[sc]_{\widetilde{A}}[\overline{sc}]_S$ tetraquark  states with the quantum numbers $J^{PC}=1^{++}$ respectively favors identifying the  $X(4140)$ and $X(4685)$   as the 1S and 2S $[sc]_S[\overline{sc}]_{A}+[sc]_{A}[\overline{sc}]_S$  or
$[sc]_S[\overline{sc}]_{\widetilde{A}}+[sc]_{\widetilde{A}}[\overline{sc}]_S$ tetraquark states respectively, see Fig.\ref{mass-1-fig} to acquire more
intuitive image. In Ref.\cite{WangZG-Di-Y4140} (also Ref.\cite{WZG-X4140-X4685}), we acquire the prediction $M_X=4.14 \pm 0.10\,\rm{GeV}$ for the 1S $[sc]_{\widetilde{V}}[\overline{sc}]_{V}-[sc]_{V}[\overline{sc}]_{\widetilde{V}}$
tetraquark state with the quantum numbers $J^{PC}=1^{++}$ in the old treating scheme, which favors  identifying the $X(4140)$ as the $[sc]_{\widetilde{V}}[\overline{sc}]_{V}-[sc]_{V}[\overline{sc}]_{\widetilde{V}}$ tetraquark state, now such an identification is superseded. In the present work based on the new treating scheme, we acquire  the prediction $M_X=4.29\pm0.09\,\rm{GeV}$ for the  1S $[sc]_{\widetilde{V}}[\overline{sc}]_{V}-[sc]_{V}[\overline{sc}]_{\widetilde{V}}$
tetraquark state, which favors identifying the $X(4274)$ as the $[sc]_{\widetilde{V}}[\overline{sc}]_{V}-[sc]_{V}[\overline{sc}]_{\widetilde{V}}$
tetraquark state with the quantum numbers $J^{PC}=1^{++}$.   In Ref.\cite{WZG-X4274-APPB},  the predictions  favor identifying the $X(4274)$ as the   $[sc]_A[\bar{s}\bar{c}]_V-[sc]_V[\bar{s}\bar{c}]_A$   tetraquark state with a relative P-wave between the diquark and antidiquark pairs, such an identification  does not suffer from shortcomings in sense of treating scheme. The $X(4274)$ maybe have two remarkable Fock components.

 We can take the pole residues $\lambda_X$ shown in Table \ref{mass-Table-ss} as the elementary parameters to investigate  the two-body
 strong decays of those hidden-charm-hidden-strange tetraquark states,
 \begin{eqnarray}
X(1^{+-}) &\to&\eta J/\psi\,  ,  \,\eta\psi^\prime\,  ,  \,\eta h_c \, ,  \, \phi \eta_c  \, , \, D_s\bar{D}_s^{*}\, ,\, D_s^{*}\bar{D}_s\, ,\, D_s^{*}\bar{D}^{*}_s\, , \nonumber\\
X(0^{++}) &\to&\eta\eta_c\,  ,  \, \eta \chi_{c1} \, ,  \, \phi J/\psi \, , \, \phi\psi^\prime \, , \,D_s\bar{D}_s\, ,\, D^{*}_s\bar{D}^{*}_s\, , \nonumber\\
X(1^{++}) &\to& \eta \chi_{c1} \, ,  \, \phi J/\psi \, , \, \phi \psi^\prime \, , \,D_s\bar{D}^{*}_s\, ,\, D^{*}_s\bar{D}_s\, ,\, D^{*}_s\bar{D}^{*}_s\, , \nonumber\\
X(2^{++}) &\to&\eta\eta_c \,  ,  \, \eta \chi_{c1}  \, ,  \, \phi J/\psi \, , \, \phi \psi^\prime \, , \,D_s\bar{D}_s\, ,\, D_s^{*}\bar{D}^{*}_s\, ,
\end{eqnarray}
with the traditional  three-point QCD sum rules or the light-cone QCD sum rules, and obtain the partial decay widths, therefore the branching  fractions or ratios,  to diagnose the nature of those $X$ states.

In 2021, the BESIII collaboration observed an new structure  near the $D_s^- D^{*0}$ and $D^{*-}_s D^0$  thresholds in the $K^{+}$-recoil mass spectrum with the significance of  5.3 $\sigma$ in the process $e^+e^-\to K^+ (D_s^- D^{*0} + D^{*-}_s D^0)$ \cite{BES3985}.
 The Breit-Wigner  mass and width are  $3985.2^{+2.1}_{-2.0}\pm1.7\,\rm{MeV}$   and $13.8^{+8.1}_{-5.2}\pm4.9\,\rm{MeV}$, respectively.

Also in 2021, the LHCb collaboration reported  two new  exotic states  in the $J/\psi  K^+$  invariant mass spectrum  in the exclusive    $B^+ \to J/\psi \phi K^+$ decays \cite{LHCb-X4685}.  The most significant state, $Z_{cs}(4000)$, has a mass of $4003 \pm 6 {}^{+4}_{-14}\,\rm{MeV}$, a width of $131 \pm 15 \pm 26\,\rm{MeV}$, and the spin-parity
$J^P =1^+$, while the broader state, $Z_{cs}(4220)$, has a mass of $4216 \pm 24{}^{ +43}_{-30}\,\rm{MeV}$, a width of $233 \pm 52 {}^{+97}_{-73}\,\rm{MeV}$ \cite{LHCb-X4685}.

As noted in our previous work \cite{WZG-Zcs3985-4110} (also in Ref.\cite{Maiani-3985-4003}),  there exist two tetraquark nonets with the symbolic  diquark structures,
\begin{eqnarray}\label{1-nonet}
 I=1 &:& [uc]_S[\bar{d}\bar{c}]_A-[uc]_A[\bar{d}\bar{c}]_S\, , \,[dc]_S[\bar{u}\bar{c}]_A-[dc]_A[\bar{u}\bar{c}]_S\, ,\, \nonumber\\ &&\frac{[uc]_S[\bar{u}\bar{c}]_A-[uc]_A[\bar{u}\bar{c}]_S-[dc]_S[\bar{d}\bar{c}]_A+[dc]_A[\bar{d}\bar{c}]_S}{\sqrt{2}}\, ; \nonumber \\
I=0 &:& \frac{[uc]_S[\bar{u}\bar{c}]_A-[uc]_A[\bar{u}\bar{c}]_S+[dc]_S[\bar{d}\bar{c}]_A-[dc]_A[\bar{d}\bar{c}]_S}{\sqrt{2}}\, ,\nonumber\\
     & &[sc]_S[\bar{s}\bar{c}]_A-[sc]_A[\bar{s}\bar{c}]_S\, ; \nonumber\\
I=\frac{1}{2} & :&[qc]_S[\bar{s}\bar{c}]_A-[qc]_A[\bar{s}\bar{c}]_S\, , \,[sc]_S[\bar{q}\bar{c}]_A-[sc]_A[\bar{q}\bar{c}]_S\, ,
\end{eqnarray}
and
\begin{eqnarray}\label{2-nonet}
 I=1 &:& [uc]_S[\bar{d}\bar{c}]_A+[uc]_A[\bar{d}\bar{c}]_S\, , \,[dc]_S[\bar{u}\bar{c}]_A+[dc]_A[\bar{u}\bar{c}]_S\, ,\, \nonumber\\ &&\frac{[uc]_S[\bar{u}\bar{c}]_A+[uc]_A[\bar{u}\bar{c}]_S-[dc]_S[\bar{d}\bar{c}]_A-[dc]_A[\bar{d}\bar{c}]_S}{\sqrt{2}} \, ;\nonumber \\
I=0 &:& \frac{[uc]_S[\bar{u}\bar{c}]_A+[uc]_A[\bar{u}\bar{c}]_S+[dc]_S[\bar{d}\bar{c}]_A+[dc]_A[\bar{d}\bar{c}]_S}{\sqrt{2}}\, ,\nonumber\\
     & &[sc]_S[\bar{s}\bar{c}]_A+[sc]_A[\bar{s}\bar{c}]_S\, ; \nonumber\\
I=\frac{1}{2} & :&[qc]_S[\bar{s}\bar{c}]_A+[qc]_A[\bar{s}\bar{c}]_S\, , \,[sc]_S[\bar{q}\bar{c}]_A+[sc]_A[\bar{q}\bar{c}]_S\, ,
\end{eqnarray}
with the charge conjugation $C=-$ and $+$, respectively, where the $I=1$, $0$, $\frac{1}{2}$ are the isospins, and $q=u$, $d$. There are enough rooms to accommodate the $X(3876)$, $Z_c(3900)$, $Z_{cs}(3985)$, $Z_{cs}(4000)$ and $X(4140)$ in a suitable way, and the $Z_{cs}(3985)$ and $Z_{cs}(4000)$ are not necessary to be the same particle.

The Table \ref{Identifications-Table} is adopted  from Ref.\cite{WZG-HC-spectrum-PRD}, where the Isospin limit is adopted, thus the tetraquark states with the symbolic quark structures $\bar{c}c\bar{d}u$, $\bar{c}c\bar{u}d$, $\bar{c}c\frac{\bar{u}u-\bar{d}d}{\sqrt{2}}$ and $\bar{c}c\frac{\bar{u}u+\bar{d}d}{\sqrt{2}}$ have the degenerated  masses.
 In Ref.\cite{WZG-HC-spectrum-PRD},  we  investigate the  mass spectrum of the ground state hidden-charm tetraquark states with the quantum numbers $J^{PC}=0^{++}$, $1^{+-}$, $1^{++}$ and $2^{++}$ in the framework of
the QCD sum rules in a comprehensive way, and  revisit the identifications of  the $X(3860)$, $X(3872)$, $X(3915)$,  $X(3940)$, $X(4160)$, $Z_c(3900)$, $Z_c(4020)$, $Z_c(4050)$, $Z_c(4055)$, $Z_c(4100)$, $Z_c(4200)$, $Z_c(4250)$, $Z_c(4430)$, $Z_c(4600)$, etc in the  scenario of tetraquark  states in a consistent/reasonable way.

Recently, the LHCb collaboration explored the decays   $B^{+}\to {D^{\ast+}D^{-}K^{+}}$ and ${D^{\ast-}D^{+}K^{+}}$, and observed four charmonium(-like) states
  $\eta_c(3945)$, $h_c(4000)$, $\chi_{c1}(4010)$ and $h_c(4300)$ with the quantum numbers $J^{PC}=0^{-+}$,  $1^{+-}$, $1^{++}$ and $1^{+-}$ respectively  in the $D^{\ast\pm}D^{\mp}$ invariant mass spectrum \cite{LHCb-hc4000}.
The measured Breit-Wigner masses and widths are
\begin{flalign}
 & \eta_c(3945) : M = 3945\,_{-17}^{+28}{}\,_{-28}^{+37} \mbox{ MeV}\, , \, \Gamma = 130\,_{-49}^{+92}{}\,_{-70}^{+101} \mbox{ MeV} \, , \nonumber\\
 & h_c(4000) : M = 4000\,_{-14}^{+17}{}\,_{-22}^{+29} \mbox{ MeV}\, , \, \Gamma = 184\,_{-45}^{+71}{}\,_{-61}^{+97} \mbox{ MeV}\, ,\nonumber \\
 & \chi_{c1}(4010) : M = 4012.5\,_{-3.9}^{+3.6}{}\,_{-3.7}^{+4.1} \mbox{ MeV} \, ,\, \Gamma = 62.7\,_{-6.4}^{+7.0}{}\,_{-6.6}^{+6.4}\mbox{ MeV} \, , \nonumber\\
 & h_c(4300) : M = 4307.3\,_{-6.6}^{+6.4}{}\,_{-4.1}^{+3.3} \mbox{ MeV} \, ,\, \Gamma = 58\,_{-16}^{+28}{}\,_{-25}^{+28} \mbox{ MeV} \, .
\end{flalign}

 In Table \ref{Identifications-Table}, there are enough rooms to accommodate the $h_c(4000)$ and $\chi_{c1}(4010)$ in scenario of tetraquark states, as the central values of the predicted  tetraquark masses happen to coincide with the experimental data. In the traditional charmonium scenario, the $h_c^\prime$ and $\chi_{c1}^\prime$ have the masses $3956\,\rm{MeV}$ and $3953\,\rm{MeV}$, respectively, there exist  discrepancies    about $50-60\,\rm{MeV}$.
The hidden-charm/hidden-charm-hidden-strange tetraquark masses acquired in our works can be confronted to the experimental data  at the BESIII, LHCb, Belle II,  CEPC, FCC, ILC  in the future to examine their nature.

\begin{table}
\begin{center}
\begin{tabular}{|c|c|c|c|c|c|c|c|c|}\hline\hline
 $X_c$              &$J^{PC}$ & $T^2 (\rm{GeV}^2)$ & $\sqrt{s_0}(\rm GeV) $      &$\mu(\rm{GeV})$   &pole         &$|D(10)|$ \\ \hline

$[sc]_{S}[\overline{sc}]_{S}$  &$0^{++}$  &$3.1-3.5$  &$4.65\pm0.10$               &$1.4$   &$(40-61)\%$  &$\ll 1\%$   \\

$[sc]_{A}[\overline{sc}]_{A}$  &$0^{++}$  &$3.1-3.5$  &$4.70\pm0.10$               &$1.5$   &$(39-60)\%$  &$\ll1\%$    \\

$[sc]_{\tilde{A}}[\overline{sc}]_{\tilde{A}}$  &$0^{++}$  &$3.4-3.9$  &$4.75\pm0.10$    &$1.6$    &$(39-61)\%$    &$\ll1\%$    \\

$[sc]_{V}[\overline{sc}]_{V}$   &$0^{++}$  &$4.0-4.5$  &$5.40\pm0.10$               &$2.8$     &$(41-60)\%$  &$\ll1\%$    \\

$[sc]_{\tilde{V}}[\overline{sc}]_{\tilde{V}}$  &$0^{++}$  &$5.2-6.1$
&$6.05\pm0.10$   &$3.7$   &$(40-61)\%$  &$\ll 1\%$   \\

$[sc]_{P}[\overline{sc}]_{P}$  &$0^{++}$  &$5.2-6.1$   &$6.10\pm0.10$               &$3.8$     &$(40-61)\%$  &$\ll1\%$    \\ \hline

$[sc]_{S}[\overline{sc}]_{S}{}^*$  &$0^{++}$  &$2.8-3.2$  &$4.50\pm0.10$               &$1.3$   &$(39-61)\%$  &$\ll 1\%$   \\

$[sc]_{A}[\overline{sc}]_{A}{}^*$  &$0^{++}$  &$2.7-3.1$  &$4.55\pm0.10$               &$1.3$   &$(39-62)\%$  &$\leq1\%$    \\

$[sc]_{\tilde{A}}[\overline{sc}]_{\tilde{A}}{}^*$  &$0^{++}$  &$3.0-3.5$  &$4.60\pm0.10$    &$1.4$    &$(39-62)\%$    &$\ll1\%$    \\

$[sc]_{V}[\overline{sc}]_{V}{}^*$   &$0^{++}$  &$3.6-4.1$  &$5.25\pm0.10$               &$2.6$     &$(40-62)\%$  &$\ll1\%$    \\

$[sc]_{\tilde{V}}[\overline{sc}]_{\tilde{V}}{}^*$  &$0^{++}$  &$4.7-5.4$
&$5.86\pm0.10$   &$3.5$   &$(41-60)\%$  &$\ll 1\%$   \\

$[sc]_{P}[\overline{sc}]_{P}{}^*$  &$0^{++}$  &$4.8-5.6$   &$5.95\pm0.10$               &$3.7$     &$(40-61)\%$  &$\ll1\%$    \\ \hline

$[sc]_S[\overline{sc}]_{A}-[sc]_{A}[\overline{sc}]_S$  &$1^{+-}$  &$3.2-3.7$          &$4.70\pm0.10$   &$1.5$  &$(39-61)\%$  &$\ll1\%$    \\

$[sc]_{A}[\overline{sc}]_{A}$     &$1^{+-}$  &$3.4-3.8$   &$4.76\pm0.10$               &$1.6$   &$(41-60)\%$    &$\ll 1\%$  \\

$[sc]_S[\overline{sc}]_{\widetilde{A}}-[sc]_{\widetilde{A}}[\overline{sc}]_S$
&$1^{+-}$ & $3.4-3.9$    &$4.76\pm0.10$   &$1.6$  &$(39-61)\%$  &$\ll 1\%$  \\

$[sc]_{\widetilde{A}}[\overline{sc}]_{A}-[sc]_{A}[\overline{sc}]_{\widetilde{A}}$
&$1^{+-}$  &$3.4-3.8$    &$4.76\pm0.10$    &$1.6$   &$(40-60)\%$ &$\ll 1\%$ \\

$[sc]_{\widetilde{V}}[\overline{sc}]_{V}+[sc]_{V}[\overline{sc}]_{\widetilde{V}}$
&$1^{+-}$  &$4.0-4.5$   &$5.40\pm0.10$  &$2.8$   &$(40-60)\%$  &$\ll 1\%$ \\

$[sc]_{V}[\overline{sc}]_{V}$  &$1^{+-}$   &$5.4-6.3$   &$6.16\pm0.10$               &$3.8$  &$(41-61)\%$  &$\ll 1\%$  \\

$[sc]_P[\overline{sc}]_{V}+[sc]_{V}[\overline{sc}]_P$   &$1^{+-}$  &$4.6-5.3$          &$5.70\pm0.10$   &$3.2$   &$(41-61)\%$  &$\ll1\%$   \\ \hline

$[sc]_S[\overline{sc}]_{A}+[sc]_{A}[\overline{sc}]_S$  &$1^{++}$   &$3.2-3.6$          &$4.70\pm0.10$    &$1.5$  &$(41-61)\%$  &$\ll 1\%$  \\

$[sc]_S[\overline{sc}]_{\widetilde{A}}+[sc]_{\widetilde{A}}[\overline{sc}]_S$
&$1^{++}$ &$3.4-3.9$    &$4.76\pm0.10$   &$1.6$   &$(39-60)\%$ &$\ll 1\%$  \\

$[sc]_{\widetilde{V}}[\overline{sc}]_{V}-[sc]_{V}[\overline{sc}]_{\widetilde{V}}$
&$1^{++}$ & $3.2-3.6$   &$4.86\pm0.10$   &$1.9$   &$(40-61)\%$ &$\ll1\%$ \\

$[sc]_{\widetilde{A}}[\overline{sc}]_{A}+[sc]_{A}[\overline{sc}]_{\widetilde{A}}$
&$1^{++}$ &$4.9-5.8$    &$5.93\pm0.10$    &$3.5$   &$(39-61)\%$ &$\ll 1\%$ \\

$[sc]_P[\overline{sc}]_{V}-[sc]_{V}[\overline{sc}]_P$  &$1^{++}$  &$4.6-5.4$          &$5.70\pm0.10$   &$3.2$   &$(39-61)\%$  &$\ll1\%$   \\ \hline

$[sc]_{A}[\overline{sc}]_{A}$    &$2^{++}$  &$3.5-4.0$   &$4.82\pm0.10$               &$1.8$   &$(40-60)\%$     &$\ll1\%$ \\

$[sc]_{V}[\overline{sc}]_{V}$    &$2^{++}$  &$5.1-6.0$   &$6.05\pm0.10$               &$3.7$     &$(40-61)\%$   &$\ll1\%$ \\
\hline\hline
\end{tabular}
\end{center}
\caption{ The Borel parameters, continuum threshold parameters, energy scales of the QCD spectral densities,  pole contributions, and contributions of the vacuum condensates of dimension $10$  for the ground state hidden-charm-hidden-strange tetraquark states. }\label{BorelP-ss}
\end{table}

\begin{table}
\begin{center}
\begin{tabular}{|c|c|c|c|c|c|c|c|c|}\hline\hline
$X_c$    &$J^{PC}$  &$M_X (\rm{GeV})$   &$\lambda_X (\rm{GeV}^5) $   \\ \hline

$[sc]_{S}[\overline{sc}]_{S}$  &$0^{++}$  &$4.08\pm0.09$  &$(3.18\pm0.52)\times 10^{-2}$  \\

$[sc]_{A}[\overline{sc}]_{A}$  &$0^{++}$  &$4.13\pm0.09$  &$(6.02\pm1.02)\times 10^{-2}$  \\

$[sc]_{\tilde{A}}[\overline{sc}]_{\tilde{A}}$  &$0^{++}$  &$4.16\pm0.09$  &$(5.83\pm0.87)\times 10^{-2}$  \\

$[sc]_{V}[\overline{sc}]_{V}$   &$0^{++}$  &$4.82\pm0.09$ &$(1.70\pm0.25)\times 10^{-1}$   \\

$[sc]_{\tilde{V}}[\overline{sc}]_{\tilde{V}}$ &$0^{++}$   &$5.46\pm0.10$
&$(5.37\pm0.58)\times 10^{-1}$  \\

$[sc]_{P}[\overline{sc}]_{P}$  &$0^{++}$  &$5.54\pm0.10$  &$(2.13\pm0.22)\times 10^{-1}$    \\ \hline

$[sc]_{S}[\overline{sc}]_{S}{}^*$  &$0^{++}$  &$3.99\pm0.09$  &$(2.41\pm0.42)\times 10^{-2}$  \\

$[sc]_{A}[\overline{sc}]_{A}{}^*$  &$0^{++}$  &$4.04\pm0.09$  &$(4.24\pm0.76)\times 10^{-2}$  \\

$[sc]_{\tilde{A}}[\overline{sc}]_{\tilde{A}}{}^*$  &$0^{++}$  &$4.08\pm0.08$  &$(4.39\pm0.69)\times 10^{-2}$  \\

$[sc]_{V}[\overline{sc}]_{V}{}^*$   &$0^{++}$  &$4.70\pm0.09$ &$(1.31\pm0.23)\times 10^{-1}$   \\

$[sc]_{\tilde{V}}[\overline{sc}]_{\tilde{V}}{}^*$ &$0^{++}$   &$5.37\pm0.11$
&$(4.41\pm0.47)\times 10^{-1}$  \\

$[sc]_{P}[\overline{sc}]_{P}{}^*$  &$0^{++}$  &$5.47\pm0.11$  &$(1.86\pm0.19)\times 10^{-1}$       \\  \hline

$[sc]_S[\overline{sc}]_{A}-[sc]_{A}[\overline{sc}]_S$   &$1^{+-}$  &$4.11\pm0.10$      &$(2.91\pm0.48)\times 10^{-2}$        \\

$[sc]_{A}[\overline{sc}]_{A}$     &$1^{+-}$  &$4.17\pm0.08$      &$(3.65\pm0.55)\times 10^{-2}$           \\

$[sc]_S[\overline{sc}]_{\widetilde{A}}-[sc]_{\widetilde{A}}[\overline{sc}]_S$
&$1^{+-}$   &$4.17\pm0.09$    &$(3.71\pm0.57)\times 10^{-2}$    \\

$[sc]_{\widetilde{A}}[\overline{sc}]_{A}-[sc]_{A}[\overline{sc}]_{\widetilde{A}}$
&$1^{+-}$   &$4.18\pm0.09$    & $(7.50\pm1.12)\times 10^{-2}$    \\

$[sc]_{\widetilde{V}}[\overline{sc}]_{V}+[sc]_{V}[\overline{sc}]_{\widetilde{V}}$
&$1^{+-}$   &$4.82\pm0.09$    &$(1.45\pm0.23)\times 10^{-1}$    \\

$[sc]_{V}[\overline{sc}]_{V}$    &$1^{+-}$  &$5.57\pm0.11$  &$(1.89\pm0.19)\times 10^{-1}$    \\

$[sc]_P[\overline{sc}]_{V}+[sc]_{V}[\overline{sc}]_P$  &$1^{+-}$  &$5.13\pm0.10$      &$(1.33\pm0.16)\times 10^{-1}$  \\ \hline

$[sc]_S[\overline{sc}]_{A}+[sc]_{A}[\overline{sc}]_S$   &$1^{++}$  &$4.11\pm0.09$      &$(2.88\pm0.46)\times 10^{-2}$     \\

$[sc]_S[\overline{sc}]_{\widetilde{A}}+[sc]_{\widetilde{A}}[\overline{sc}]_S$
 &$1^{++}$   &$4.17\pm0.09$    &$(3.67\pm0.57)\times 10^{-2}$    \\

$[sc]_{\widetilde{V}}[\overline{sc}]_{V}-[sc]_{V}[\overline{sc}]_{\widetilde{V}}$
&$1^{++}$   &$4.29\pm0.09$    &$(5.49\pm0.92)\times 10^{-2}$    \\

$[sc]_{\widetilde{A}}[\overline{sc}]_{A}+[sc]_{A}[\overline{sc}]_{\widetilde{A}}$
&$1^{++}$   &$5.34\pm0.10$    &$(2.52\pm0.30)\times 10^{-1}$    \\

$[sc]_P[\overline{sc}]_{V}-[sc]_{V}[\overline{sc}]_P$  &$1^{++}$  &$5.12\pm0.10$      &$(1.33\pm0.17)\times 10^{-1}$    \\ \hline

$[sc]_{A}[\overline{sc}]_{A}$   &$2^{++}$  &$4.24\pm0.09$   &$(6.03\pm0.88)\times 10^{-2}$      \\

$[sc]_{V}[\overline{sc}]_{V}$    &$2^{++}$  &$5.49\pm0.11$   &$(2.43\pm0.26)\times 10^{-1}$      \\
\hline\hline
\end{tabular}
\end{center}
\caption{ The masses and pole residues of the ground state hidden-charm-hidden-strange  tetraquark states. }\label{mass-Table-ss}
\end{table}

\begin{table}
\begin{center}
\begin{tabular}{|c|c|c|c|c|c|c|c|c|}\hline\hline
$X_c$      &$J^{PC}$  & $M_X(\rm{GeV})$   & Identifications    &$X_c^\prime$ \\ \hline

$[sc]_{S}[\overline{sc}]_{S}$  &$0^{++}$  &$4.08\pm0.09$   &     & \\

$[sc]_{A}[\overline{sc}]_{A}$  &$0^{++}$  &$4.13\pm0.09$   &       &?\,$X(4700)$ \\

$[sc]_{\tilde{A}}[\overline{sc}]_{\tilde{A}}$  &$0^{++}$  &$4.16\pm0.09$  &                     &?\,$X(4700)$ \\

$[sc]_{V}[\overline{sc}]_{V}$  &$0^{++}$  &$4.82\pm0.09$  &    & \\

$[sc]_{\tilde{V}}[\overline{sc}]_{\tilde{V}}$  &$0^{++}$  &$5.46\pm0.10$  &                   &  \\

$[sc]_{P}[\overline{sc}]_{P}$  &$0^{++}$  &$5.54\pm0.10$     &  &  \\ \hline

$[sc]_{S}[\overline{sc}]_{S}{}^*$  &$0^{++}$  &$3.99\pm0.09$   &?\,$X(3960)$      &?\,$X(4500)$   \\

$[sc]_{A}[\overline{sc}]_{A}{}^*$  &$0^{++}$  &$4.04\pm0.09$   &     & \\

$[sc]_{\tilde{A}}[\overline{sc}]_{\tilde{A}}{}^*$  &$0^{++}$  &$4.08\pm0.08$  &                    & \\

$[sc]_{V}[\overline{sc}]_{V}{}^*$  &$0^{++}$  &$4.70\pm0.09$  &?\,$X(4700)$  & \\

$[sc]_{\tilde{V}}[\overline{sc}]_{\tilde{V}}{}^*$  &$0^{++}$  &$5.37\pm0.11$  &                   &  \\

$[sc]_{P}[\overline{sc}]_{P}{}^*$  &$0^{++}$  &$5.47\pm0.11$     &  &  \\ \hline

$[sc]_S[\overline{sc}]_{A}-[sc]_{A}[\overline{sc}]_S$  &$1^{+-}$  &$4.11\pm0.10$      &       &   \\

$[sc]_{A}[\overline{sc}]_{A}$     &$1^{+-}$  &$4.17\pm0.08$    & &   \\

$[sc]_S[\overline{sc}]_{\widetilde{A}}-[sc]_{\widetilde{A}}[\overline{sc}]_S$
 &$1^{+-}$   &$4.17\pm0.09$   &  &    \\

$[sc]_{\widetilde{A}}[\overline{sc}]_{A}-[sc]_{A}[\overline{sc}]_{\widetilde{A}}$
&$1^{+-}$   &$4.18\pm0.09$    &  &   \\

$[sc]_{\widetilde{V}}[\overline{sc}]_{V}+[sc]_{V}[\overline{sc}]_{\widetilde{V}}$
&$1^{+-}$   &$4.82\pm0.09$    &        &    \\

$[sc]_{V}[\overline{sc}]_{V}$   &$1^{+-}$  &$5.57\pm0.11$   &   &  \\

$[sc]_P[\overline{sc}]_{V}+[sc]_{V}[\overline{sc}]_P$   &$1^{+-}$  &$5.13\pm0.10$        &   &  \\ \hline

$[sc]_S[\overline{sc}]_{A}+[sc]_{A}[\overline{sc}]_S$   &$1^{++}$  &$4.11\pm0.09$     &?\,$X(4140)$       &?\,$X(4685)$   \\

$[sc]_S[\overline{sc}]_{\widetilde{A}}+[sc]_{\widetilde{A}}[\overline{sc}]_S$
 &$1^{++}$   &$4.17\pm0.09$    &?\,$X(4140)$  &?\,$X(4685)$   \\

$[sc]_{\widetilde{V}}[\overline{sc}]_{V}-[sc]_{V}[\overline{sc}]_{\widetilde{V}}$
&$1^{++}$   &$4.29\pm0.09$    &?\,$X(4274)$ &    \\

$[sc]_{\widetilde{A}}[\overline{sc}]_{A}+[sc]_{A}[\overline{sc}]_{\widetilde{A}}$
&$1^{++}$   &$5.34\pm0.10$    &      & \\

$[sc]_P[\overline{sc}]_{V}-[sc]_{V}[\overline{sc}]_P$  &$1^{++}$  &$5.12\pm0.10$      &   &  \\ \hline

$[sc]_{A}[\overline{sc}]_{A}$   &$2^{++}$  &$4.24\pm0.09$     &  & \\

$[sc]_{V}[\overline{sc}]_{V}$   &$2^{++}$  &$5.49\pm0.11$      &   & \\
\hline\hline
\end{tabular}
\end{center}
\caption{ The probable identifications of the hidden-charm-hidden-strange tetraquark states. }\label{Identifications-Table-ss}
\end{table}

\begin{table}
\begin{center}
\begin{tabular}{|c|c|c|c|c|c|c|c|c|}\hline\hline
$Z_c$($X_c$)                                                            & $J^{PC}$  & $M_{X/Z}(\rm{GeV})$   & Identifications     &$Z_c^\prime$ ($X_c^\prime$)      \\ \hline

$[uc]_{S}[\overline{dc}]_{S}$                                           & $0^{++}$  & $3.88\pm0.09$      & ?\,$X(3860)$       &       \\

$[uc]_{A}[\overline{dc}]_{A}$                                           & $0^{++}$  & $3.95\pm0.09$      & ?\,$X(3915)$       & \\

$[uc]_{\tilde{A}}[\overline{dc}]_{\tilde{A}}$                           & $0^{++}$  & $3.98\pm0.08$      &                    & \\

$[uc]_{V}[\overline{dc}]_{V}$                                           & $0^{++}$  & $4.65\pm0.09$      &                    & \\

$[uc]_{\tilde{V}}[\overline{dc}]_{\tilde{V}}$                           & $0^{++}$  & $5.35\pm0.09$      &                    &  \\

$[uc]_{P}[\overline{dc}]_{P}$                                           & $0^{++}$  & $5.49\pm0.09$      &                    &  \\ \hline

$[uc]_S[\overline{dc}]_{A}-[uc]_{A}[\overline{dc}]_S$                   & $1^{+-}$  & $3.90\pm0.08$      & ?\,$Z_c(3900)$      &?\,$Z_c(4430)$    \\

$[uc]_{A}[\overline{dc}]_{A}$                                           & $1^{+-}$  & $4.02\pm0.09$      & ?\,$Z_c(4020/4055)$ &?\,$Z_c(4600)$        \\

$[uc]_S[\overline{dc}]_{\widetilde{A}}-[uc]_{\widetilde{A}}[\overline{dc}]_S$     & $1^{+-}$   & $4.01\pm0.09$    & ?\,$Z_c(4020/4055)$ &?\,$Z_c(4600)$     \\

$[uc]_{\widetilde{A}}[\overline{dc}]_{A}-[uc]_{A}[\overline{dc}]_{\widetilde{A}}$ & $1^{+-}$   & $4.02\pm0.09$    & ?\,$Z_c(4020/4055)$ &?\,$Z_c(4600)$    \\

$[uc]_{\widetilde{V}}[\overline{dc}]_{V}+[uc]_{V}[\overline{dc}]_{\widetilde{V}}$ & $1^{+-}$   & $4.66\pm0.10$    & ?\,$Z_c(4600)$      &    \\

$[uc]_{V}[\overline{dc}]_{V}$                                           & $1^{+-}$  & $5.46\pm0.09$      &                    &  \\

$[uc]_P[\overline{dc}]_{V}+[uc]_{V}[\overline{dc}]_P$                   & $1^{+-}$  & $5.45\pm0.09$      &                    &  \\
\hline

$[uc]_S[\overline{dc}]_{A}+[uc]_{A}[\overline{dc}]_S$                   & $1^{++}$  & $3.91\pm0.08$      & ?\,$X(3872)$       &   \\

$[uc]_S[\overline{dc}]_{\widetilde{A}}+[uc]_{\widetilde{A}}[\overline{dc}]_S$     & $1^{++}$   & $4.02\pm0.09$    &?\,$Z_c(4050)$ &   \\

$[uc]_{\widetilde{V}}[\overline{dc}]_{V}-[uc]_{V}[\overline{dc}]_{\widetilde{V}}$ & $1^{++}$   & $4.08\pm0.09$    &?\,$Z_c(4050)$ &    \\

$[uc]_{\widetilde{A}}[\overline{dc}]_{A}+[uc]_{A}[\overline{dc}]_{\widetilde{A}}$ & $1^{++}$   & $5.19\pm0.09$    &               & \\

$[uc]_P[\overline{dc}]_{V}-[uc]_{V}[\overline{dc}]_P$                   & $1^{++}$  & $5.46\pm0.09$      &                    &  \\
\hline

$[uc]_{A}[\overline{dc}]_{A}$                                           & $2^{++}$  & $4.08\pm0.09$      &?\,$Z_c(4050)$      & \\

$[uc]_{V}[\overline{dc}]_{V}$                                           & $2^{++}$ & $5.40\pm0.09$      &                    & \\ \hline \hline

$[uc]_{A}[\overline{dc}]_{A}$                                           & $1^{+-}$  & $4.02\pm0.09$      & ?\,$h_c(4000)$ &        \\

$[uc]_S[\overline{dc}]_{\widetilde{A}}-[uc]_{\widetilde{A}}[\overline{dc}]_S$     & $1^{+-}$   & $4.01\pm0.09$    & ?\,$h_c(4000)$ &      \\

$[uc]_{\widetilde{A}}[\overline{dc}]_{A}-[uc]_{A}[\overline{dc}]_{\widetilde{A}}$ & $1^{+-}$   & $4.02\pm0.09$    & ?\,$h_c(4000)$ &    \\

$[uc]_S[\overline{dc}]_{\widetilde{A}}+[uc]_{\widetilde{A}}[\overline{dc}]_S$     & $1^{++}$   & $4.02\pm0.09$    &?\,$\chi_{c1}(4010)$ &   \\

\hline\hline
\end{tabular}
\end{center}
\caption{ The probable  identifications of the  hidden-charm tetraquark states, the isospin limit is implied. This Table is adopted from Ref.\cite{WZG-HC-spectrum-PRD} and new identifications are added. }\label{Identifications-Table}
\end{table}

\section{Conclusion}
In the present work,  we construct the diquark-antidiquark type four-quark currents to investigate  the ground state hidden-charm-hidden-strange  tetraquark states with the quantum numbers $J^{PC}=0^{++}$, $1^{+-}$, $1^{++}$ and $2^{++}$ at length in the framework of
the traditional QCD sum rules in a comprehensive way. In analytical computations,   we accomplish the operator product expansion up to the vacuum condensates of dimension $10$ using the counting rules and vacuum saturation in a rigorous way. We  update old calculations and perform new calculations, therefore we acquire concise and elegant expressions, and  could perform  whole analysis in a perfectly consistent way.  Then we  apply  the modified energy scale formula $\mu=\sqrt{M^2_{X/Y/Z}-(2{\mathbb{M}}_c)^2}-2{\mathbb{M}}_s$ to select the suitable energy scales to extract the hadronic parameters so as to take  the collective/net light-flavor $SU(3)$ breaking effects into account in a perfectly consistent way.
The present predictions favor identifying the $X(3960)$ and $X(4500)$ as the 1S and 2S tetraquark states with the quantum numbers $J^{PC}=0^{++}$ respectively; identifying the  $X(4700)$ as the first radial excitation of the $[sc]_{A}[\overline{sc}]_{A}$ or $[sc]_{\tilde{A}}[\overline{sc}]_{\tilde{A}}$ tetraquark state with the quantum numbers $J^{PC}=0^{++}$;
 identifying the $X(4700)$ as the $[sc]_{V}[\overline{sc}]_{V}{}^*$ tetraquark state with the quantum numbers $J^{PC}=0^{++}$;
  identifying the  $X(4140)$ and $X(4685)$   as the 1S and 2S $[sc]_S[\overline{sc}]_{A}+[sc]_{A}[\overline{sc}]_S$  or
$[sc]_S[\overline{sc}]_{\widetilde{A}}+[sc]_{\widetilde{A}}[\overline{sc}]_S$ tetraquark states with the quantum numbers $J^{PC}=1^{++}$ respectively; identifying the $X(4274)$ as the $[sc]_{\widetilde{V}}[\overline{sc}]_{V}-[sc]_{V}[\overline{sc}]_{\widetilde{V}}$
tetraquark state with the quantum numbers $J^{PC}=1^{++}$.
We should admit that we still lack enough experimental data and theoretical works to make  unambiguous identifications.
Furthermore, we consider our previous theoretical  predictions, and try to make possible identifications of the new LHCb exotic states, more precisely,  we tentatively identify the  $h_c(4000)$ as the $[uc]_{A}[\overline{dc}]_{A}$, $[uc]_S[\overline{dc}]_{\widetilde{A}}-[uc]_{\widetilde{A}}[\overline{dc}]_S$ or                                             $[uc]_{\widetilde{A}}[\overline{dc}]_{A}-[uc]_{A}[\overline{dc}]_{\widetilde{A}}$ tetraquark state with the quantum numbers $J^{PC}=1^{+-}$, and identify the $\chi_{c1}(4010)$ as
the
$[uc]_S[\overline{dc}]_{\widetilde{A}}+[uc]_{\widetilde{A}}[\overline{dc}]_S$ tetraquark state with the   quantum numbers    $J^{PC}=1^{++}$.
The predicted tetraquark states can be confronted to the experimental data in the future at the BESIII, LHCb, Belle II,  CEPC, FCC, ILC to examine the exotic states.

\section*{Acknowledgements}
This  work is supported by National Natural Science Foundation, Grant Number  12175068, and Hebei Key Laboratory of Physics and Energy Technology.

\end{document}